\documentclass[aps,prd,showpacs,onecolumn,nofootinbib,superscriptaddress,amsmath,amssymb,amsthm]{revtex4}

\usepackage{epsfig}
\usepackage{graphicx}

\usepackage{subcaption}

\usepackage{bm,dsfont,braket,mathrsfs}
\usepackage{color}

\usepackage{definitions}%

\newcommand{\CB}{{\cal B}}
\newcommand{\CC}{{\cal C}}

\newtheorem{defi}{Definition}
\newtheorem{theor}{Theorem}

\begin{document}
\title{Precise Wigner-Weyl calculus for lattice models}

\author{I.V. Fialkovsky}
\email{ifialk@gmail.com}
\affiliation{Physics Department, Ariel University, Ariel 40700, Israel}
\affiliation{CMCC-Universidade  Federal  do  ABC,  Santo  Andre,  S.P.,  Brazil}


\author{M.A. Zubkov \footnote{On leave of absence from Institute for Theoretical and Experimental Physics, B. Cheremushkinskaya 25, Moscow, 117259, Russia}}
\email{zubkov@itep.ru}
\affiliation{Physics Department, Ariel University, Ariel 40700, Israel}

\date{\today}

\begin{abstract}
We propose a new version of Wigner-Weyl calculus for tight-binding lattice models. It allows to express various physical quantities through Weyl symbols of operators and Green's functions. In particular, Hall conductivity in the presence of varying and arbitrarily strong magnetic field is represented using the proposed formalism as a topological invariant.
\end{abstract}

\pacs{73.43.-f}

\maketitle
\tableofcontents

\section{Introduction}

Topology plays an important role both in relativistic quantum field theory and in condensed matter physics. One of its most well-known manifestations is nowadays associated with the Quantum Hall Effect (QHE). The first topological representation of the Hall conductivity has been proposed for homogeneous systems in the presence of constant magnetic field. It is called the TKNN invariant \cite{Thouless1982}.  The TKNN invariant is also used for description of homogeneous Hall insulators that possess intrinsic anomalous quantum Hall effect (AQHE) in absence of external fields. The TKNN invariant is given by an integral of Berry curvature over the occupied energy levels.
The corresponding value is not changed when the dependence of energy on momentum varies smoothly, see  \cite{Avron1983,Fradkin1991,Hatsugai1997,Qi2008} and reviews \cite{Kaufmann:2015lga,Tong:2016kpv}.

Momentum space topological invariants were widely studied within the context of condensed matter physics since then. First of all, in the absence of interactions the Hall conductivity for the intrinsic AQHE has been expressed through the Green's function \cite{IshikawaMatsuyama1986,Volovik1988} (see also Chapter 21.2.1 in \cite{Volovik2003a}). For two-dimensional topological insulators the Hall conductivity is given by
$$
\sigma_H = \frac{\cal N}{2\pi},
$$
where
{\be
{\cal N}
=  \frac{ \epsilon_{ijk}}{  \,3!\,4\pi^2}\, \int d^3p \Tr
\[
{G}(p ) \frac{\partial {G}^{-1}(p )}{\partial p_i}  \frac{\partial  {G}(p )}{\partial p_j}  \frac{\partial  {G}^{-1}(p )}{\partial p_k}
\].
\label{N-0}
\ee}%
It appeared later  that the simplest possible topological invariant in momentum space is composed of the two-point Green's  function of a homogeneous fermionic model, and  is responsible for the stability of the Fermi surface \cite{Volovik2003a}
\be
N_1= \tr \oint_C \frac{dp^l}{2\pi \ii}
G(p_0,p)\partial_l G^{-1}(p_0,p).
\label{N1}
\ee
Here $\partial^\rho \equiv \partial/\partial p_\rho$, while $C$ is a closed path, which encloses the Fermi surface in momentum space. The topological stability of  Fermi points is protected by another invariant, which reveals correspondence with the above mentioned expression for the Hall conductivity  \cite{Matsuyama1987a,Volovik2003a}
\be
N_3=\frac1{24\pi^2} \epsilon_{\mu\nu\rho\lambda} \tr\int_S dS^\mu
G\partial^\nu G^{-1}
G\partial^\rho G^{-1}
G\partial^\lambda G^{-1}.
\label{N3}
\ee
Here $S$ is a surface surrounding the Fermi point. These and more involved constructions are widely used in condensed matter physics theory \cite{HasanKane2010,Xiao-LiangQi2011,Volovik2011,Volovik2007,VolovikSemimetal}. Topological invariants in momentum space are responsible for the gapless nature of  fermions at the edges of topological insulators \cite{Gurarie2011a,EssinGurarie2011} and in the bulk of Weyl semi-metals \cite{Volovik2003a,VolovikSemimetal}. The fermion zero modes are related to various topological defects in $^3$He-superfluid, and are protected by the similar topological invariants \cite{Volovik2016}. In the high energy theory the topology of momentum space has been discussed, for example, in \cite{NielsenNinomiya1981a,NielsenNinomiya1981b,So1985a,IshikawaMatsuyama1986,Kaplan1992a,Golterman1993,Volovik2003a,Hovrava2005,Creutz2008a,Kaplan2011}.

By construction, the TKNN invariant, as well as (\ref{N-0}-\ref{N3}), is defined for systems without interactions. It allows to deal with {physically homogeneous configurations even if the gauge potential corresponding to the constant magnetic field is coordinate dependent. In the latter case the magnetic Brillouin zone is to be introduced in order to define the TKNN invariant.} The important question is what will happen to the topological representation of Hall conductivity if the system becomes non-homogeneous and interacting. The inhomogeneity can be given by the electric field of impurities, elastic deformations, or by the variations of magnetic field. Up to now the representation of Hall conductivity in the general case as a topological invariant  has not been constructed. However, several particular cases have been investigated.

Although TKNN and \Ref{N-0} were obtained for the case when there are no interactions, according to the common lore they both remain valid for the models with weak interactions, if the two-point Green's  function is taken with the interaction corrections. This statement was proved rigourously in $2+1$D relativistic Quantum Electrodynamics \cite{ColemanHill1985,Lee1986}.  Recently \cite{ZZ2019} the AQHE has been considered for the models of rather general types. There the mentioned statement has been proven, at least, in one-loop approximation.

For the systems with homogeneous magnetic field the influence of interactions on the Hall conductivity has been discussed widely (see, for example \cite{KuboHasegawa1959,Niu1985a,Altshuler0,Altshuler} and references therein). The general output of these studies is that the weak interactions do not affect the value of Hall conductivity. The case of varying magnetic field has been considered recently in the series of papers written with the participation of the authors on the present paper.
{
	In \cite{ZW2019} a new expression for the Hall conductivity has been proposed. It was given in  a topologically invariant form, composed, however, of approximate Wigner transformation of the  two-point Green's  functions.
	The adaptation of Weyl symbol used in \cite{ZW2019} is described in Appendix \ref{app:conti_W} of the present paper.}
Graphene in the presence of inhomogeneities invoked by elastic deformations was considered in \cite{FZ2019}. In \cite{ZZ2019_2} the proof was presented that in the presence of interactions the Hall conductivity is still given by the expression proposed in \cite{ZW2019} with the complete two-point Green's  function. The mentioned version of the Wigner-Weyl calculus has been also discussed in \cite{Suleymanov2019}. This approximate formalism allows for calculation of the Hall conductivity in the case when the magnitude of magnetic field is much smaller than several thousands Tesla, while the wavelengths are much larger than $1$ Angstrom. Under these conditions the average Hall conductivity of a two dimensional systems is given by $
\sigma_H = \frac{\cal N}{2\pi},
$ with
\be
{\cal N}
=  \frac{T \epsilon_{ijk}}{ |{\bf A}| \,3!\,4\pi^2}\, \int \D{{}^3x} \int_{\cM}  \D{{}^3p}
	\, {\rm tr}\, {G}_{\cC}(x,p )\star \frac{\partial {Q}_{\cC}(x,p )}{\partial p_i} \star \frac{\partial  {G}_{\cC}(x,p )}{\partial p_j} \star \frac{\partial  {Q}_{\cC}(x,p )}{\partial p_k}
\label{calM2d230I}
\ee
Here $T$ is temperature that is assumed to be small, $|{\bf A}| $ is the (infinite) area of the system, ${G}_{\cC}(x,p )$ is a type of Wigner transformation (see Eq. \ref{GWxAH}) of the two-point Green's function $\hat G$, being in its turn the inverse of the Dirac operator $\hat{Q}$. As it was mentioned above those quantities are defined using the specific version of the Wigner-Weyl calculus designed especially for the consideration of the QHE {in weak magnetic fields. In  general case, it gives only an approximate expression for the conductivity.}

The original Wigner-Weyl calculus stemmed from the works of  H. Groenewold \cite{Groenewold1946} and J. Moyal \cite{Moyal1949}, who considered the alternative formulations of quantum mechanics in infinite continuous space. This calculus accumulated the ideas of H. Weyl \cite{Weyl1927} and E. Wigner \cite{Wigner1932}. According to this formalism it is possible to use the Wigner distribution instead of the wave function, Weyl symbol instead of the operators of the observable quantities, and Moyal product in lieu of operator one \cite{Ali2005,Berezin1972}. The original version of Wigner-Weyl calculus has been widely applied to quantum mechanics \cite{Curtright2012,Zachos2005}. Certain modifications of Wigner-Weyl formalism were proposed within the context of both high energy field theory and condensed matter physics \cite{Cohen1966,Agarwal1970,E.C.1963,Glauber1963,Husimi1940,Cahill1969,Buot2009}.
The Wigner distribution has been used in QCD \cite{Lorce2011,Elze1986}. It has been considered also in quantum kinetic theory \cite{Hebenstreit2010,Calzetta1988}, in and in the noncommutative field theories \cite{Bastos2008,Dayi2002}. There were several attempts to apply various versions of the Wigner-Weyl calculus in a variety of physical problems including cosmology \cite{Habib1990,Chapman1994,Berry1977}.

{Formulation of the Wigner-Weyl calculus  on a lattice (or on a torus) faced certain difficulties, and still this field of research is not settled. It started from the works of Schwinger \cite{Schwinger570}, and continued throughout the century. We would like to mention the works of F.Buot \cite{Buot1974,Buot2009,Buot2013}, Wooters \cite{WOOTTERS19871}, and Leonhardt \cite{Leonhardt1995} devoted to the physical applications, and the works of Kasperowitz \cite{KASPERKOVITZ199421}, and Ligab\'o \cite{Ligabo2016} on the mathematical formulations of the discreet Wigner-Weyl calculus.  The closely related area of research is the deformational quantization \cite{BAYEN197861,Kontsevich2003}.  Several important general facts were proven in this field, including the existence of a trace operation \cite{Felder2000}, and perturbative construction of the Moyal product \cite{Kupriyanov2008}. However, the mentioned results are either too abstract or are possessing certain caveats, impeding the real physical applications. }

The approximate version of the lattice Wigner-Weyl calculus, based on $G_\cC$ from above is described in Appendices \ref{app:conti_W} and \ref{app:old_W} of the present paper.
The application of this calculus, as already mentioned, is limited to slowly varying weak fields. However, even in this form it appears to be powerful enough to describe response of various nondissipative currents to external field strength \cite{Zubkov2017,Chernodub2017,Khaidukov2017,Zubkov2018a,Zubkov2016a,Zubkov2016b}. In many cases such a response is expressed through topological invariants, which do not changed when the given model is modified smoothly. In this way the absence of the equilibrium chiral magnetic effect \cite{Kharzeev2014} has been proven \cite{Zubkov2016a}, the AQHE has been investigated \cite{Zubkov2016b}, the chiral separation effect (CSE) has been reestablished \cite{Metlitski2005} within the properly regularized models (using lattice field theory) \cite{Khaidukov2017,Zubkov2017}. This formalism has also been applied to the  high density QCD \cite{Zubkov2018a}, and to the field systems in the presence of both gravity and magnetic field (the so-called scale magnetic effect) \cite{Chernodub2016,Chernodub2017}.

{
In the present paper we consummate the development of the Wigner-Weyl calculus for lattice models, aiming primarily at the systems in external magnetic field.
We present the {\it precise} version of this calculus, which allows us to obtain topological expression for the Hall conductivity for the models defined on rectangular lattices without any limitations on the magnitude of external magnetic field and on the rate of its variations as a function of coordinates.

In the next Section we formulate our main result (upon introducing all necessary notation) in the form of four Theorems, and prove them one by one in the consequent Sections \ref{sec:WW-QFT} and \ref{sec:W-symb}.
}

\section{Main result}

Before the formulation of our main result let us introduce notations and definitions used throughout the paper. We assume the relativistic units with $c = \hbar=1$, if not explicitly stated otherwise.

\subsection{The Hilbert space}
\label{H-space}

In the one-dimensional models the physical lattice is denoted by $\cO$, and for simplicity of the forthcoming considerations we write it as
\be
\cO= \{2\ell k, k\in \dZ\}.
\ee
The first Brillouin zone is
\be
\cM=(-\pi/(2\ell),\pi/(2\ell)].
\ee
Apart from that, we also consider an extended lattice $\fO$
\be
\fO\equiv \{ \ell k, k\in \dZ\}=\cO\cup\cO', \qquad
\cO'= \{\ell (2k+1), k\in \dZ\}
\label{fO}
\ee
Its first Brillouin zone is $\fM = (-\pi/\ell,\pi/\ell]$. The additional lattice $\cO'$ can be considered as a translation of the physical one
\be
\cO'=\cO+\ell .
\ee

The usual properties of the physical states are assumed
\be
\hat 1_\cO=\sum_{x\in\cO} \ket{x}\bra{x}= \int_{\cM}\D{p} \ket{p}\bra{p},\quad
\braket{x|p}=\frac1{\sqrt{|\cM|}} e^{\ii x p},\quad
\braket{p|q} = \delta(p-q\bmod{\pi/\ell}), \quad \braket{x|y}= \delta_{x,y}.
\label{ord-states}
\ee
The latter is a Kronecker symbol since our coordinate is discreet. Accordingly, we have the Fourier decomposition
\be
\ket{p}
= \frac1{\sqrt{|\cM|}}\sum_{x \in \cO}e^{\ii p x}\ket{x}.
\label{ket bp0}
\ee

In the present paper we will also consider a $D$-dimensional rectangular lattice. It will still be denoted by $\cO$. Its points and the corresponding Brillouin zone are
\be
	\cO=\left\{\sum_{j=1}^D 2 k_j  \belj,\ k_j\in\dZ\right\},\quad
	\cM = \left\{\sum_{j=1}^D \al_j \bgj/2,\ \al_j \in(-1/2,1/2]\right\},
\ee
$$
\belj \bel^{(i)} = (\ell_j)^2 \delta_{ij},\quad
\belj\bg^{(i)}=2\pi \delta_{ij},\quad i,j=1,2,\ldots D.
$$
We assume that the theory to be dealt with is Euclidean equilibrium one. For brevity we assume the discretization of imaginary time together with the discretization of space coordinates. However, in the applications to condensed matter physics typically imaginary time is continuous. Therefore, speaking of certain properties of our constructions we will take off the discretization of imaginary time. This will be pointed out explicitly. If the absence of the discretization of time is not mentioned explicitly, such a discretization is implied.

The immediate generalization of \Ref{ord-states} and \Ref{ket bp0} to multi-dimensional case is assumed to hold.
The auxiliary extended lattice $\fO$ in $D$-dimensional case, and the corresponding extended momentum space $\fM$ are given by:
\be
	\fO=\left\{\sum_{j=1}^D k_j \belj,\ k_j\in\dZ\right\},\quad
	\fM = \left\{\sum_{j=1}^D \al_j \bgj,\ \al_j\in(-1/2,1/2]\right\},\qquad
	\belj\bg^{(i)}=2\pi \delta_{ij},\quad i,j=1,2,\ldots D.
\ee
To avoid the clatter of notation, we will not use bold-face notation in what follows.
The physical Hilbert space $\cH$ is defined using the basis $|x\rangle$ ($x\in \cO$) that obeys Eq. (\ref{ord-states}). All operators to be discussed below are assumed to be defined on this Hilbert space.

\subsection{Lattice Weyl symbol}\label{SectAxiom}

An abstract lattice Weyl symbol is defined in the following way.
\begin{defi}\label{w-def}
	By the Weyl symbol of a linear operator $\hat A$ acting on the Hilbert space $\cH$ we understand the map $\hat A\in \mathscr{L}(\cH)\mapsto A_W(x,p)\in L^2(\cM\times\fO)$ such that
	there could be defined a $\star$-product between Weyl symbols of any two operators, as well as trace operation $\rm Tr$ mapping $L^2(\cM\times\fO)$ to complex numbers, $\mathds C$, satisfying the following conditions
	\begin{enumerate}
		\item{Star product identity}
		\be
		A_W(x,p) \star B_W(x,p) = (\hat A\hat B)_W(x,p).
		\label{*-def}
		\ee
		\item{First trace identity}
		\be
		\Tr {A_W} = \tr \hat A
		\label{Tr-def-1}
		\ee
		\item{Second trace identity}
		\be
		\Tr [A_W(x,p) \star B_W(x,p)] = \Tr [A_W(x,p) B_W(x,p)].
		\label{Tr-def-2}
		\ee
		\item{Weyl symbol of identity operator}
		\be
		({\hat 1})_W(x,p) = 1.
		\label{id-def}
		\ee
	\end{enumerate}
By $\rm tr$ we understand the trace of the operator itself in the original Hilbert space.
\end{defi}
An explicit construction of such Weyl symbol leads to
\begin{theor}\label{th-Q-W}
	The integral representation
	\be
		Q_W(x,p) =  \int_{\cM} \D{q} e^{2 \ii q x}
		Q(p+q,p-q) \prod_{j=1}^D\(1+e^{-2 \ii q\ell^{(j)}}\)
		\label{Q-Weyl-fO}
	\ee
	is a Weyl symbol of operator $\hat Q$ in the sense of the  Definition above, while
	\be
		 Q(p,q) = \frac{1}{|\cM|}\sum_{x_1,x_2\in \cO}\bra{x_1} \hat Q \ket{x_2}
		e^{\ii(x_2 q -x_1 p) }.
		\label{Q-Fourier}
	\ee
\end{theor}
In this case
\be
\Tr Q_W
	\equiv
	\Tr_\fO Q_W
	= \frac{1}{|\fM|}\int_{\cM} \D{p}\sum_{x\in \fO}\tr Q_W(x,p).
\label{Tr_W}
\ee
{Here $\tr$ stands for the trace over the inner symmetries (if any), which we will omit in the future.} The $\star$-product is the original Moyal one
\be
\star
	\equiv\star_{x,p}
	=e^{-\frac\ii2\vec\partial_x \cev\partial_p+\frac\ii2\cev\partial_x\vec\partial_p}.
\label{*-W}
\ee
This pseudo-differential operator acting on functions of discreet coordinate $x$ is understood either as an integral operator, see \Ref{A*B-1}, or as acting on analytical continuation, see the end of the Section.

We will show that the integral representation \Ref{Q-Weyl-fO} is equivalent to a series representation for the Weyl symbol, which can be formulated as
\begin{theor}\label{th-Q-series}
	For operators, possessing a series representation of a function of two variables
	\be
		\hat{Q} = \cQ(x,p)\big|_{x\to \ii\partial_p} =\sum_n (\ii\partial_p)^n Q_n(p) \label{Q1}
	\ee
	its Weyl symbol can be defined as
	\be
		{Q}_{\cal W}\(x,p\) = \sum_n x^n q_{n}(p),
		\label{Q-Weyl-ser}
	\ee
	with
	\be
		q_n(p) = \sum_{k\ge n}
		C_{k}^{n} \Big(\frac{\ii}{2}\partial_p\Big)^{k-n} Q_n(p).
		\label{Q3}
	\ee
	Moreover,
	\be
	Q_\cW\(x,p\) = Q_W\(x,p\),\quad \forall x\in\fO
	\ee
	provided that
	\be
	\cQ(x+\ell, p) = \cQ(x, p), \quad \forall x\in\cO.
	\ee
\end{theor}
The notation above is evident in one-dimensional case. In $D$ dimensions, the indices are to treated as multi-indices, $n = (n_1, ..., n_D)$, etc., with $ (\ii\partial_p)^n = \prod_i (\ii\partial_{p_{i}})^{n_i}$ and $C_{k}^{n}=\prod_i C_{k_i}^{n_i}$. A sum over a multi-index is understand as a sum over all ot its components, and a condition of the type of $n < k$ where $n$ and $k$ are the multi-indices, implies that the sum is over all those values of components that obey $n_i < k_i$ for any $i$.

Altogether, throughout the paper we will use three different symbols of operators: $Q_W$ given by \Ref{Q-Weyl-fO}, $Q_\cW$ of \Ref{Q-Weyl-ser}, and also the symbol of operator proposed by Buot \cite{Buot1974},
\be
	Q_\cB (x,p) = \int_{\cM}  \D{q} e^{2\ii x q} Q({p+q},{p-q}).
	\label{Q-Buot}
\ee
The relation between the former two is given by the theorem above, while the latter does not satisfy our conditions for a proper Weyl symbol. However, it will be useful for establishing the properties of $Q_W$ and $Q_\cW$. Buot's symbol is investigated in details in Appendix \ref{app:buot}.

We work with the model defined on the lattice. Therefore, the application of the derivatives in $x$ present in the star product is questionable at a first look. The problem is that the analytical continuation of function defined on discreet (countable) set of points is not unique, generally speaking, \cite{Rubel1956}. However, it appears that the application of the $\star$-product to the symbols of operators does not depend on the choice of the analytical continuation, provided that we know their values on the extended lattice $\fM$.

Let us consider this problem for simplicity for one-dimensional lattice. Generalization to the multidimensional case is evident. The symbols we deal with ($Q_\cW$, $Q_W$ and $Q_\cB$) are all $\pi/\ell$-periodic in $p$. Thus they can be represented as
\be
Q_{\cW / W/ \cB}(x,p) = \sum _{n\in\dZ} e^{2 \ii p n \ell} q_n(x)
\ee
with certain functions $q_n(x)$ defined on the lattice $x\in\fM = \ell \dZ$. Constructing an analytical continuation
\be
  q_n(z):   q_n(\ell \dZ) = q_n(\ell \dZ)
\label{anali}
\ee
we investigate the $\star$-product of two such objects
\be
Q_W*S_W
=\sum_{n,m\in\dZ}  e^{2 \ii p n \ell}  e^{2 \ii p m \ell}   q_n(x+m\ell)    s_m(x-n\ell)\label{starproduct}
\ee
to see that it does not depend on the choice of the analytical continuation in \Ref{anali}, as the star acts by shifting the arguments of $q_n$ and $s_m$ in such a way that those arguments always belong to $\fO$. So, the application of $\star$-product is actually independent of the particularities of the continuation requiring only the knowledge of the function's  values on $x\in\fO$.

\subsection{Global current and averaged conductivity}
\label{maincond}
In Euclidian space-time the partition function of a physical system (defined by its Dirac operator $\hat Q$) is given by
\be
Z = \int D\bar{\psi }D\psi
\,\, e^{S[\psi ,\bar\psi  ]}
\label{Z01}
\ee
with the action
\be
S[\psi ,\bar\psi  ]
	\equiv \braket {\psi |\hat Q |\psi }
	= \int \D{p dq} \,
\bar\psi  (p)\, Q(p,q)\,\psi (q),
\label{S}
\ee
where the integration measure and normalization are understood to be chosen appropriately for the model under consideration. Note, that $\bar\psi$ stands for the second independent integration Grassman variable, and has nothing to do with the Dirac spinor conjugated operator, used in the operator formalism.

Using \Ref{w-def} and {\it Definition} \ref{w-def}, the action can be written as
\be
S[\psi ,\bar\psi  ]
	= \Tr \(\rho_W(x,p) Q_W(x,p)\),
\ee
where
$\rho_W(x,p)$ is the Weyl symbol of an operator $\hat{\rho}$ with Grassman-valued matrix elements $\rho(x,y) \equiv \bar{\psi}(x) \psi(y) = \langle x | \hat{\rho} | y \rangle$
\be
\rho_W(x,p) = \( \hat{\rho} \)_W.
\ee
Then, we can formulate and prove the following two theorems
\begin{theor}\label{th-current}
	In the system described by \Ref{S} the total electric current of the sample is given by
	\be
		\bar{J_k} = -\Tr( G_W \star \partial_{p_k} Q_W),
	\ee
	where
	$$
	G_W(x,p) = (\hat Q^{-1})_W(x,p).
	$$
	Moreover, this expression is a topological invariant: it is not changed under small variations of the lattice Dirac operator $\hat Q$.
\end{theor}
\begin{theor}\label{th-conduct}
	In the system described by \Ref{S} subject to  constant external electromagnetic field $F^{(E)}_{lm}$, the total electric current averaged over the volume $|{\bf V}|$ of the sample is given by
	\be
	{\mathcal J}_k
	= {\mathcal W}_{lmk}F^{(E)}_{lm } ,
	\label{cJ}
	\ee
	\be
	{\mathcal W}_{lmk}
	\equiv \frac{\ii  }{2{\beta |{\bf V}|}}
	\Tr \(
	G_W^{(0)} \star \partial_{p_l} Q_W^{(0)} \star
	G_W^{(0)} \star \partial_{p_m} Q_W^{(0)} \star
	G_W^{(0)} \star \partial_{p_k} Q_W^{(0)}
	\).
	\label{cW}
	\ee
	
	The Hall conductivity, in turn,
	\be
	\bar\sigma_{mk} \equiv 	({\mathcal W}_{Dmk}-	{\mathcal W}_{Dkm})/i,
	\label{bar_s}
	\ee
	is a topological invariant.
\end{theor}
Above $\beta = 1/T$ is the inverse temperature, $T$ is assumed to be small. We also assume that the system is gapped, so that the integral in Eq. (\ref{cW}) does not contain divergencies.

\section{Wigner-Weyl Quantum Field Theory}
\label{sec:WW-QFT}

In this section we describe an arbitrary non-interacting fermionic model defined on the lattice. The basic quantities are represented in terms of the Weyl symbols of the corresponding operators. We assume here, that there exists a Weyl symbol obeying the conditions given in the above {\it Definition} \ref{w-def}. Using these properties (but not the particular form of the Weyl symbol) we will derive in this section the expression for the quantum Hall conductivity. It is expressed through the Weyl symbol of the fermion propagator, and appears to be a topological invariant as soon as its expression does not contain divergences (in practice this occurs if the system is gapped, and the Fermi energy lies in the gap). Thus in this section we will prove {\it Theorem} \ref{th-current} and {\it Theorem} \ref{th-conduct}. The explicit construction of the Weyl symbol of operator (and the proof of {\it Theorem} \ref{th-Q-W} and {\it Theorem}  \ref{th-Q-series}) will be given in the next Section.

\subsection{Lattice model in momentum space}

\label{SectLatMod}

Following \cite{Zubkov2016a,Zubkov2016b,Suleymanov2019,ZW2019} we start with the lattice tight-binding fermionic model with the partition function of the following form
\begin{equation}
	Z=\int D\bar\psi   D\psi   \, \exp\left(\sum_{x,y}\bar\psi  (x)\left(-i\mathcal{D}_{x,y}\right)\psi (y)\right)\,.
	\label{Z00}
\end{equation}
Here $\mathcal{D}_{x,y}$ is a matrix that depends on the discrete lattice coordinates $x, y \in \cO$. $\psi ,\bar\psi  $ are multi-component Grassmann-valued fields defined on the lattice sites. The corresponding indices are omitted here and below for brevity. Normalization of the functional integration measure is assumed to be chosen  appropriately for the problem at hand.

{In the simplest case of a uniform system} such a partition function can be rewritten in momentum space as follows
\be
Z
	= \int D\bar\psi  D\psi \, {\rm exp}\(  \int_{\cal M} \frac{d^D {p}}{|{\cal M}|}\bar\psi ({p}){Q}(p)\psi({p}) \)\,,\label{Z01}
\ee
where integration is over the fields defined in momentum space $\cal M$. $|{\cal M}|$  is its volume, $D$ is the dimensionality of space-time (we will omit it whenever this does not lead to misunderstanding), $\bar{\psi}(p)$ and $\psi(p)$ are the anti-commuting multi-component Grassmann variables, now defined in momentum space. Without loss of generality we assume that time is discretized, so that momentum space is compact, and its volume is finite. In condensed matter physics the imaginary time typically is not discretized,  the corresponding partition function can be obtained easily as the limit of Eq. \Ref{Z00} when the (imaginary) time spacing tends to zero. The partition function of Eq. (\ref{Z01}) allows to describe non-interacting fermionic systems corresponding to matrix  $Q(p)$ (that is the Fourier transform of the lattice tight-binding operator $D_{x,y}$). The meaning of $Q(p)$ for the lattice models of electrons in crystals is the inverse propagator of Bloch electron.

Introduction of an external gauge field $A(x)$ defined as a function of coordinates effectively leads to the Peierls substitution (see, for example, \cite{Zubkov2016a,Zubkov2016b,Suleymanov2019}):
\be
Z
	= \int D\bar{\psi}D\psi \, {\rm exp}\Big(  \int_{\cal M} \frac{d  {p}}{|{\cal M}|} \bar\psi  ({p}){Q}(p - A(\ii\partial_p))\psi({p}) \Big),\label{Z01a}
\ee
where the products of operators inside expression ${Q}(p - A(\ii\partial_p))$ are symmetrized.

We relate operator $\hat{Q} = Q(p-A(\ii\partial_p))$ and its inverse $\hat{G} = \hat{Q}^{-1}$ defined in Hilbert space ${\cal H}$ of functions (on $\cal M$) with their matrix elements ${ Q}(p,q)$ and ${ G}(p,q)$ correspondingly, in the usual way
$$
{Q}(p,q) = \braket{ p|\hat{Q}| q}, \quad {G}(p,q) = \braket{ p|\hat{Q}^{-1}| q}\,.
$$
Here the basis elements of $\cal H$ are normalized as in \Ref{ord-states}. These operators obey the following equation:
$$
\braket{ p|\hat{Q}\hat{G}|q} = \delta({p} - {q})\,.
$$

{In the non-uniform case (either due to the external gauge potential $A(x)$, or because of any other reason)} Eq. (\ref{Z01a}) can be rewritten as
\be
Z= \int D\bar\psi  D\psi \,
	\exp\(
		\int_{\cal M} \frac{d  {p}_1}{\sqrt{|{\cal M}|}} \int_{\cal M} \frac{d  {p}_2}{\sqrt{|{\cal M}|}} \bar\psi ({p}_1){Q}(p_1,p_2)\psi({p}_2)
	\),
	\label{Z1}
\ee
while the Green function of Bloch electron is given by
\be
{G}_{ab}(k_2,k_1)
	= \frac{1}{Z}\int D\bar{\psi}D\psi \, \exp\(  \int_{\cal M} \frac{d  {p}_1}{\sqrt{|{\cal M}|}} \int_{\cal M} \frac{d  {p}_2}{\sqrt{|{\cal M}|}}\bar{\psi}({p}_1){Q}(p_1,p_2)\psi({p}_2) \) \frac{\bar{\psi}_b(k_2)}{\sqrt{|{\cal M}|}} \frac{\psi_a(k_1)}{\sqrt{|{\cal M}|}}\label{G1}\,.
\ee
Here indices $a,b$ enumerate the components of the fermionic fields. In the following we will omit those indices for brevity.

\subsection{Electric current in the Wigner-Weyl formalism}

Let us suppose, that we modified the external gauge field as ${A} \rightarrow {A} + \delta { A}$. The original external field $A$ can vary arbitrarily, but $\delta A$ is assumed to be slowly varying, i.e. its variation at the distance of the lattice spacing can be neglected. In the linear response theory, the functional derivative of partition function with respect to this extra contribution to the gauge potential gives the electric current. In the tight-binding model we can write,
$$
\delta \log Z = \sum_{[xy]} j_{[xy]} \int_x^y \delta A(u)du.
$$
Here the sum is over all lattice links $[xy]$, while $j_{[xy]}$  is the current along the link $[xy]$. Since the latter is constant along a link and $\delta A$ can be considered constant as well, we can rewrite the above expression as
\be
\delta \log Z
	 = \sum_{x\in\cO} j_{\mu}(x) \delta A_\mu(x)
	 = \frac1{2^D}\sum_{x\in\fO} j_{\mu}(x) \delta A_\mu(x).
\label{dj}
\ee
The former sum is over the physical points $\cO$, while the latter is over the extended lattice $\fO$, where there are $2^D$ more points than in  $\cO$.

For the variation of the partition function we have
\be
\begin{split}
{\delta} \, {\rm log}\, Z
&=
	\frac{1}{Z} \int D{\bar \psi} D\psi \, {\rm exp}\(  \int_{\cal M} \frac{d  { p}}{|{\cal M}|} \bar{\psi}({ p})\hat{Q}(\ii{\partial}_{ p},{ p})\psi ({ p}) \) \, \int_{\cal M} \frac{d  { p}}{|{\cal M}|} \bar{\psi }({ p})\Big[\delta \hat{Q}(\ii{\partial}_{ p},{ p})\Big]\psi ({ p}) \\
& =
	\int_{\cal M} {\D  { p}}  \delta \hat{Q}(\ii{\partial}_{{ p}_1},{ p}_1)\,G({ p}_1,{ p}_2) \Big|_{{ p}_1 = { p}_2 = { p}}\\
& =
	\int_{\cal M} {\D  { p}}  \braket{ p| \hat{G} \delta \hat{Q} |p}
	\equiv\tr \(\hat{G} \delta \hat{Q}\).
	\label{j4}
\end{split}
\ee
The trace over the fermionic indices (and any inner symmetries) is implied here.

The definition and the properties of the Weyl symbols allow us to rewrite the last expression as a trace of the Weyl symbols of corresponding operators
\be
{\delta} \, {\rm log}\, Z
	 = \Tr (\hat{G} \delta \hat{Q})_W
	 =	
	 \Tr  \[ {G}_{W} \delta {Q}_{W} \] .
	 \label{j428}
\ee
Now we use that $\delta A(x)$ (unlike $A(x)$) varies slowly at the distances of the order of the lattice spacing. This permits us to represent
\be
	\delta Q_{W}(x,p) \approx  -\partial_{p_k} Q_{W}(x,p) \delta A_k(x).
\label{delta Q}
\ee
Substituting this to \Ref{j428}, recalling the definition of $\Tr$ of \Ref{Tr_W}, and comparing with \Ref{dj}, we come to the expression for the electric current
\be
\begin{split}
j_k(x)
&=
	-
	\int_{\cM} \frac{dp}{|{\cM}|} \,  {G}_{W}(x,p) \frac{\partial}{ \partial p_k}Q_{W}(x,p) .
	\label{Jappr}
\end{split}
\ee

Local current \Ref{Jappr} is not a topological invariant. Let us calculate the total current
\be
\bar{J_k}
	\equiv \sum_{x\in\fO} j_k(x)
	= -\Tr( G_W \star \partial_{p_k} Q_W).
	\label{bar J}
\ee
It is worth mentioning, that this definition of the total current differs somehow from the conventional definition (which is the integral over the surface of the current density across the given surface). The quantity called in the present paper for brevity {\it the total current} according to Eq. (\ref{bar J}) is the sum over the whole lattice (including the imaginary time direction) of the current density. The averaging in time is of no effect, since we assume the system to be time independent.

Unlike \Ref{Jappr}, $\bar{J_k}$ is a topological invariant.  Indeed, under small variations of the Weyl symbol of the lattice Dirac operator, $ Q_W\approx Q_W + \delta Q_W$, the Green's function varies accordingly, $G_W\approx G_W +   \delta G_{W}$, and then
\be
\delta\( \Tr\[G_W \star \partial_{p_k} Q_W \] \)
=  \Tr\[G_W \star  \partial_{p_k} \delta Q_W +\delta G_{W} \star\partial_{p_k} Q_W  \].
\label{delta J-1}
\ee
Eq. \ref{id-def} guarantees that the symbol of the $\hQ \hat G =\hat 1$ becomes the simplest Groenewold equation $G_W\star Q_W=1$. Then, $ \delta G_{W}  = -G _W\star \delta Q_W \star G_W  $,  the two terms of \Ref{delta J-1} become
\be
\begin{split}
	\Tr&\[ G_W \star  \partial_{p_k} \delta Q_W-G _W*\delta Q_W*G
	\star\partial_{p_k}  Q_W  \] \\
	& = \Tr\[ G _W\star \partial_{p_k} Q_W  \star G _W\star \delta Q_W-G _W
	\star \delta Q_W\star G _W\star\partial_{p_k} Q_W  \] 	\nonumber\\	
\end{split}
\ee
where we integrated by parts and used that $ \partial_{p_l}G_{W}  = -G_W \star \partial_{p_l} Q_W  \star G_W $. Now, simple cyclic transformation inside the trace shows that
\be
	\delta \bar{J_k} =0,
\label{deltaJ}
\ee
providing for the proof of {\it Theorem}  \ref{th-current}.

{A note on the allowed variations of the Dirac operator are in order. When varying the current we are only allowed to introduce such modifications of the Dirac operator (and its Green's function) that do not break the mere existence of the integration over momenta and coordinates in \Ref{bar J}. Physically it means, that any variations of the system that give birth to zero (delocalized, propagating) modes are forbidden. These modes would naturally cause a pole in $\hat G$ and render the integration divergent. Such modifications of the system correspond to the topological phase transitions, where the invariant does indeed change (or may change, in principle) its value.
}

\subsection{Calculation of Hall conductance}
\label{sigmaTop}

Let us start from Eq. (\ref{Jappr}) for the electric current. We represent the electromagnetic potential as a sum of two contributions:
$$
	A = A^{(M)} + A^{(E)},
$$
where $A^{(E)}$ is responsible for the electric field, while $A^{(M)}$ -- for magnetic one. The former is assumed to be weak, and we expand \Ref{Jappr} up to the terms linear in $A^{(E)}$ and its derivatives.

Following \Ref{delta Q} the symbol of Dirac operator acquires the form
\be
Q_W
	\approx  Q_W^{(0)}-  \partial_{p_m} Q_W^{(0)} A^{(E)}_m .
\label{Qappr}
\ee
The Groenewold equation for $G_{W}$ then can be solved iteratively. We will keep in this solution the terms linear in $A^{(E)}$ and in its first derivative. The zeroth order term (that does not contain $A^{(E)}$ at all) is denoted $G_{W}^{(0)}$. Then
\be
G_{W}\approx G_{W}^{(0)}+  G_{W}^{(0)}\star (\partial_{p_m} Q_{W} A_m) \star G_{W}^{(0)}.
\ee
Further expanding the stars in the above expression, which contains the derivatives in $x$ acting on $A^{(E)}$, we have
\be
G_W\approx G_W^{(0)}+  G_{W,m}^{(1)} A^{(E)}_m + 	G_{W,lm}^{(2)} \partial_l A^{(E)}_m,	
\label{Gappr}
\ee
where
$$
G_{W,m}^{(1)} = G_W^{(0)}\star \partial_{p_m} Q_W^{(0)} \star G_W^{(0)},\qquad
G_{W,lm}^{(2)} = \frac\ii2 G_W^{(0)}\star \partial_{p_l} Q_W^{(0)} \star   G_W^{(0)}
\star \partial_{p_m} Q_W^{(0)} \star   G_W^{(0)}.
$$
Upon substitution of \Ref{Qappr} and \Ref{Gappr} into \Ref{Jappr} we obtain
\be
{j_k(x)}
\approx \frac{\ii  F^{(E)}_{lm }(x)}2 \int dp
\,\tr \(
G_W^{(0)} \star \partial_{p_l} Q_W^{(0)} \star
G_W^{(0)} \star \partial_{p_m} Q_W^{(0)} \star
G_W^{(0)} \boldsymbol{\cdot} \partial_{p_k} Q_W^{(0)}
\),
\label{J(x)}			
\ee
where the last product is the ordinary one. Notice that $Q_W^{(0)}=Q_W^{(0)}(x,p)$,  $G_W^{(0)}=G_W^{(0)}(x,p)$. Assuming that the external field is constant across the system, $F^{(E)}_{lm}=const$ (equivalent to neglecting higher derivatives of $A^{(E)}$, which is already done), one can calculate the total current averaged over the volume  $|{\bf V}|$ (and also averaged in time)
\be
{\mathcal J}_k \equiv \frac1{\beta |{\bf V}|}\int dx {j_k(x)}
= {\mathcal W}_{lmk}F^{(E)}_{lm } ,
\label{cJ}
\ee
\be
{\mathcal W}_{lmk}
\equiv \frac{\ii  }{2{\beta |{\bf V}|}} \Tr
	\[
		G_W^{(0)} \star \partial_{p_l} Q_W^{(0)} \star
		G_W^{(0)} \star \partial_{p_m} Q_W^{(0)} \star
		G_W^{(0)} \star \partial_{p_k} Q_W^{(0)}
	\].
\label{cW}
\ee
Here $\beta = 1/T$ is the inverse temperature, $T$ is assumed to be small. In the above expression we restored the $\star$--product in the last factor using once again \Ref{Tr-def-2}. In what follows we will omit the superscript ${}^{(0)}$ for brevity.

The averaged Hall conductivity is given now by
\be
\bar\sigma_{mk} ={\mathcal W}_{D[mk]}/i
	\equiv 	({\mathcal W}_{Dmk}-	{\mathcal W}_{Dkm})/i,
\label{bar_s}
\ee
while the other transport coefficients are given by similar expressions ${\mathcal W}_{i[mk]} $, $i=1,2$, $m,k=1,2,3$. All of them are invariant under such variations of $\hQ$ that do not break the applicability of the $\cW$, see discussion in the previous section.

Indeed, under $\hQ\to\hQ+\delta \hQ$, we can write
\be
\begin{split}
\delta\cW_{i_1[i_2 i_3]}
&
	= \frac{\ii  \epsilon_{i_1 i_2 i_3}}{2{\beta |{\bf V}|}}
	\epsilon_{abc} \Tr
	\[
	(\delta  G_W \star \partial_{i_a}Q_W + G_W \star \partial_{i_a} \delta Q_W)\star G_W \star
		\partial_{i_b} Q_W \star G_W \star \partial_{i_c} Q_W
	\]
\end{split}
\ee	
using now that $\delta G_W=-G_W \star\delta Q_W\star G_W$ and $\partial_p G_W=-G_W\star \partial_p Q_W\star G_W$
we have
\be
\begin{split}
\delta\cW_{i_1[i_2 i_3]}
&	= \frac{\ii \epsilon_{i_1 i_2 i_3} }{2{\beta |{\bf V}|}}
	\epsilon_{abc} \Tr
	\[
	\partial_{i_a}(\delta Q_W \star G_W)\star
		\partial_{i_b} Q_W  \star \partial_{i_c} G_W
	\]
\end{split}
\ee
which after the integration by parts is identically zero, finalizing the proof of Theorem \ref{th-conduct}.

For the two-dimensional system ($D=2+1$) we come to the following representation of the average Hall current (i.e. the Hall current integrated over the whole area of the sample divided by this area $|{\bf A}|$) in the presence of electric field along the $ x_2$ axis (we take into account that the corresponding Euclidean field strength has the components $F^{(E)}_{Dk} = -i E_k$):
$$
J_1 = \frac{|{\cal V}|}{2^D }\sum_{x \in \fO} j_1(R)  = \frac{\cal N}{2\pi} E_2
$$
with
\be
{\cal N}
=  \frac{T \epsilon_{ijk}}{ |{\bf A}| \,3!\,4\pi^2}\, \frac{|{\cal V}|}{2^D }\sum_{x \in \fO}\int_{\cM}  {d^3p}
	\[
	{G}_{W}(x,p )\star \frac{\partial {Q}_{W}(x,p )}{\partial p_i} \star \frac{\partial  {G}_{W}(x,p )}{\partial p_j} \star \frac{\partial  {Q}_{W}(x,p )}{\partial p_k}
	\]_{A^{(E)}=0}
\label{calM2d230}
\ee
Here $x = (R_1,R_2,\tau)$,  $|{\bf A}|$ is the area of the system, $T$ is temperature (that is assumed to be small). It is implied that the area $|{\bf A}|$ is much larger than the area of the elementary crystal lattice cell, so that we still can deal with continuous values of momenta. For the same reason the sum over the Matsubara frequencies $\omega = 2\pi(N+1/2) T$ ($N\in \dZ$) can be substituted by an integral at $T\to 0$.  Recall, that from the very beginning we discretized imaginary time, so that the whole lattice is composed of the $(D-1)$-dimensional crystal lattice and the discretized imaginary time (as it occurs in the lattice regularization of the field-theoretical models). In condensed matter physics time typically remains continuous, the Matsubara frequencies belong to the interval $(-\infty, +\infty)$ (at $T\to 0$). In this case Eq. (\ref{calM2d230}) reads:
\be
{\cal N}
=  \frac{T \epsilon_{ijk}}{ |{\bf A}| \,3!\,4\pi^2}\, \frac{|{\cal V}^{(2)}|}{2^{2} }\sum_{\vec x \in \fO}\int_0^{1/T} d\tau \int_{\cM} {d^3p} \Tr
\[
{G}_{W}(x,p )\star \frac{\partial {Q}_{W}(x,p )}{\partial p_i} \star \frac{\partial  {G}_{W}(x,p )}{\partial p_j} \star \frac{\partial  {Q}_{W}(x,p )}{\partial p_k}
\]_{A^{(E)}=0}.
\label{calM2d230}
\ee
Here $x = (\tau, \vec{x})$, $\vec{x}$ is the point in space, $\tau$ is imaginary time that varies between $0$ and $1/T \to \infty$, ${\cal V}^{(2)}$ is the area of the two - dimensional lattice cell.

The consideration of the three-dimensional systems with $D=3+1$ is completely similar. It gives the current density integrated over the whole volume divided by this volume and averaged with respect to time) in the presence of external electric field $E_i$:
\begin{eqnarray}
\braket{ j^{k}}
& =&
	\frac{1}{2\pi^2}\epsilon^{kjl4} {\cal N}_{l} E_{j} \label{calM0C},\\
{\cal N}_l
& = &
	- \frac{T\epsilon_{ijkl}}{|{\bf V}|\, 3!\,8\pi^2}\ \frac{|{\cal V}^{(3)}|}{2^{3} }\sum_{\vec x \in \fO}\int_0^{1/T} d \tau \int_{\cM} d^4p \, \Big[  {G}_W(p,x) \star \frac{\partial {Q}_W(p,x)} {\partial p_i}\star \frac{\partial  {G}_W(p,x)}{\partial p_j} \star \frac{\partial  {Q}_W(p,x)}{\partial p_k} \Big]
\end{eqnarray}
Here $|{\bf V}|$ is the overall volume, $|{\cal V}^{(3)}|$ is the volume of the three-dimensional lattice cell. Notice, that in the above expressions we deal with an equilibrium system. Therefore, function $G_W(p,x)$ entering the above expressions does not depend on imaginary time $\tau$.

\section{Weyl symbol derivation}
\label{sec:W-symb}

\subsection{Lattice Wigner-Weyl calculus through the series expansion}

The previous definitions of the Wigner-Weyl formalism (used earlier for the description of the quantum Hall effect), \cite{Zubkov2016a,Zubkov2016b}, given for completeness in Appendix \ref{app:conti_W} and Appendix \ref{app:old_W}, are sufficient for the description of the systems in the presence of slowly varying fields. This assumes, in particular the requirement for the real crystals that the magnetic field is much smaller than about $10000$ Tesla. The definition of Appendix \ref{app:buot}, originating from \cite{Buot1974}, in turn, although is valid for any magnitudes of external fields, does not satisfy the requirements needed to express the Hall conductivity through the topological invariant composed of the Weyl symbol of $\hQ$.

In order to calculate Hall conductivity using the Wigner-Weyl formalism for the systems of general type (in particular, for those that are subject to strong magnetic fields) we need the precise version of the formalism, which satisfies the above given {\it Definition} 1  unlike the Buot's version (see also Eq. \ref{starexact0}). We will modify the versions of Wigher-Weyl calculus of Appendices \ref{app:buot}-\ref{app:old_W}. We will see, that the conditions of {\it Definition} 1 (and thus the precise Groenewold equation) hold true in this case, while the Weyl symbols of simple operators remain non-degenerate.

\subsubsection{Formal Definition}

Following Appendix \ref{app:old_W} we define implicitly the symbol $Q_\cW$ of an operator $\hat Q$, but unlike the previous works \cite{Zubkov2016a,Zubkov2016b} we apply this definition both to the Dirac operators and to their Green's functions.

Namely, if an operator $\hat Q$ is given as an operator-valued function of operators $p, \ii \partial_p$:
$$
\hQ = \cQ(\ii\partial_p,p),
$$
its symbol (which will be shown to be a Weyl symbol in the sense of {\it Definition} \ref{w-def}) is given implicitly by relation
\be
\begin{split}
\int_{\fM} & \D{PdK}
	f(P,K) {Q}_{\cal W}(-\ii\cev{\partial}_{K}+\ii\vec{\partial}_{P},{P}/2+{K}/{2})\, h(P,K)
	\\
	& =  \int_{\fM}  \D{P dK} f(P,K) \cQ\(\ii \partial_{P}+ \ii\partial_{K},{P}/2+ {K}/{2}\) \, h({P},{K}), \label{corrl2}
\end{split}
\ee
which must hold true for arbitrary functions $f({P},{K})$ and $h({P},{K})$ defined on momentum space, ${P},{K}\in \fM$. The derivatives  $\vec{\partial}_{P}$ and $\cev{\partial}_{K}$ inside the arguments of $Q_\cW$ act only outside of this function, i.e. $\cev{\partial}_{K}$ acts on $f({P},{K})$ while $\vec{\partial}_{P}$  acts on $h({P},{K})$. At the same time
the derivatives without arrows act as usual operators, i.e. not only right to the function $\hat{\cal Q}$, but inside it as well. Recall that the argument $p$ of function $\cQ(x,p)$ belongs to $\cM$, while for $\cQ\(\ii \partial_{P}+ \ii\partial_{K},{P}/2+ {K}/{2}\)$ the  arguments $P,K$ belong to ${\fM}$. This is the reason why the integrals in Eq. (\ref{corrl2}) are over $\fM$.

In the following we restrict ourselves to $D$-dimensional rectangular lattices only, bet generalization to triclinic ones are immediate.
For the practical calculation of $Q_{\cal W}$ we should represent
$\hat Q$ using the Taylor epxansion for the function $\cQ(x,p)$ in vicinity of $x=0$
\be
	\cQ(x,p) =\sum_n x^n Q_n(p),\label{Q00}
\ee
then\footnote{{Notice, that this expression actually defines the operator ordering we use to relate operator $\hQ$ and function $\cQ$.}}
\be
\hQ
	= \sum_n (\ii\partial_p)^n Q_n(p)
	. \label{Q1}
\ee
Here $n = (n_1, ..., n_D)$ is a multi-index, while $ (\ii\partial_p)^n = \prod_i (\ii\partial_{p_{i}})^{n_i}$. The representation of Eq. (\ref{Q00}) as a series in powers of $x$ defines the function of $x$ in a certain (real valued) vicinity of $x=0$. For the values of $x\in \fO$, when the series in Eq. (\ref{Q00}) do not converge, we define the function $Q(x,p)$ as an analytical continuation.

For the further use, we might also write the operator as
\be
\hQ
 	= \sum_n  {Q}_n(p) (\ii\partial_p)^n
 	\label{Q1a}
\ee
with
$$
 {Q}_n(p)
	= \sum_{m_i\ge n_i \forall i}\prod_i\[ C_{m_i}^{n_i}(\ii\partial_{p_i})^{m_i-n_i}\]
	Q_{m}(p)
	\equiv \sum_{m\ge n}C_m^n (\ii\partial_p)^{m-n}Q_m(p) .
$$
{Here and below the sums are understood for every component of the multi-indices.}
It is worth mentioning that for the given operator $\hat Q$ its representation in form of the series of Eq. (\ref{Q1}) is not unique. There may exist different sets of functions $\{Q_n(p) |n=0,1,...\}$) that correspond to the same operator $\hat{Q}$ defined on the Hilbert space for the given lattice $\cO$. (That would not be so if we consider operators in continuous theories.)

In order to use \Ref{corrl2} we represent
$$
\cQ\Big(\ii \partial_{P}+ \ii\partial_{K},{P}/2+ {K}/{2}\Big)
	= \sum_{m,l}(\ii\partial_{K})^m \mathfrak{q}_{ml}(P/2+K/2) (\ii\partial_{P})^l,
$$
where
$$
\mathfrak{q}_{m l}(p)
	= \sum_{n\ge m+l}\frac{n!}{m!l!(n-m-l)!}
	\[(\ii\partial_p/2)^{n-m-l}Q_n(p)\].
$$
The original operator-valued function $\cQ$ is periodic in $p$, the same can be assumed of $Q_n(p)$ and $\mathfrak{q}_{lm}$.

Substituting the above expressions into the RHS of \Ref{corrl2} we are able to perform there the integration by parts in $K$, obtaining
\be
\begin{split}
	\int_{\fM} &\D{P d K}
	f(P,K) {Q}_\cW(-\ii \cev{\partial}_{K}+\ii\vec{\partial}_{P},{P}/2+{K}/{2})\, h(P,K)\\
	& =
	\int_{\fM} \D{P d K} f(P,K)  \(\sum_{m,l}
	(-\ii\cev\partial_{K})^m \mathfrak{q}_{ml}(P/2+K/2) (\ii\vec\partial_{P})^l \) \, h({P},{K}).
\end{split}
\label{Q_C-def}
\ee
It is solved by
$$
{Q}_{\cal W}\(x+y,p+q\)
	= \sum_{m,l}x^m \mathfrak{q}_{ml}(p+q)y^l,
$$
here $m$, $n$ are the multi-indices, $x^n=\prod_i \(x_i\)^{n_i} $. Alternatively the solution can be written as
$$
{Q}_\cW\(x,p\)
= \sum_{m,l} \frac{1}{2^{m+l}}x^{m+l} \mathfrak{q}_{ml}(p)
= \sum_{n} x^{n} \mathfrak{q}_{n0}(p).
$$
The later equality is true since
\be
\mathfrak{q}_{n0}(p)
= \frac{1}{2^n}\sum_{m=0}^n \mathfrak{q}_{m,n-m}(p).
\ee

Therefore,
\be
{Q}_{\cal W}\(x,p\) = \sum_n x^n \mathfrak{q}_{n}(p),
\label{Q2}
\ee
where
\be
\mathfrak{q}_{n}(p)
= \sum_{k\ge n} C_{k}^{n} \[\({\ii}\partial_p/2\)^{k-n} Q_n(p)\].
\label{Q3}
\ee
The given set of functions $\{Q_n(p)|n=0,1,...\}$ defines uniquely the operator $\hat Q$. However, as it was mentioned above, the given operator $\hat Q$ does not define uniquely the set of functions $\{Q_n(p)|n=0,1,...\}$. Correspondingly, the above expressions Eq. (\ref{Q2}) and (\ref{Q3}) define a set of different symbols of an operator. Further we will formulate the restrictions on this set to be used for the construction of an appropriate unique Weyl symbol.

\subsubsection{Integral representation for $Q_{\cal W}$}
\label{QWI}

Our next purpose is derivation of an  integral representation for $Q_{\cal W}$. First of all, for $x\in \fO$ we can represent it as follows:
\be
\begin{split}
Q_{\cal W}(x,p)
&
	= \int_{\cal M} \D{k} e^{2\ii  x k} {Q}_{\cal W}(-i\cev{\partial}_k/2 + i\vec{\partial}_p/2,p+k)\delta^{[\pi/\ell]}(k)\\
&
	= \int_{\cal M} \D{k} e^{2\ii  x k} \cQ(\ii\partial_p/2 + i\partial_k/2,p+k)\delta^{[\pi/\ell]}(k)\\
&
	= 2^D \int_{\cal M} \D{k} e^{2\ii  x k} \cQ (\ii\partial_p/2 + i\partial_k/2,p+k)\braket{p+k|p-k}\prod_{i=1}^D\frac{1+e^{2\ii k_i\ell_i}}{2}.
	\label{QcW_int1}
\end{split}
\ee
Here we defined the $D$-dimensional periodic delta function as
\be
\delta^{[2\pi/\ell]}(k)
	\equiv  \prod_{i=1}^D \sum_{n_i\in\dZ} \delta(k_i-2\pi n_i/\ell_i),
\ee
and used the following relations
\be
\delta^{[\pi/\ell]}(k)
= {2^D} \delta^{[2\pi/\ell]}(2k)
=
 {2^D} \delta^{[\pi/\ell]}(2k)\prod_{i=1}^D \frac{1+e^{2\ii k_i\ell_i}}{2}.
\ee
Recall, that $\braket{p+k|p-k} = \delta^{[\pi/\ell]}(2 k)$, see \Ref{ord-states}.

Notice that
\be
\prod_{i=1}^D({1+e^{2\ii k_i\ell_i}})
	= \sum_{U} \prod_{i\in U} e^{2\ii k_i\ell_i},
\ee
where the sum is over all sets $U$, $U\subset \{1,2,\ldots D\}$. We plug it into the integrand of \Ref{QcW_int1} and denoting, for brevity, $p_\pm = p\pm k$, we represent each term of the sum over $U$ using \Ref{Q1} as:
\be
\begin{split}
\cQ(\ii\partial_{p_+},p_+)
& 	\braket{p_+|p_-}\prod_{i\in U} e^{\ii (p_{+,i}-p_{-,i})\ell_i}
	= \sum_n {Q}_n(p_+) (\ii\partial_{p_+})^n \braket{ p_+|p_-} \prod_{i\in U} e^{\ii (p_{+,i}-p_{-,i})\ell_i} \\
&	=
	\sum_n  {Q}_n(p_+) \sum_{0\le k  \le n}C_n^k
	\Big((\ii\partial_{p_+})^{k} \braket{ p_+|p_-}\Big) \Big((\ii\partial_{p_+})^{n-k} \prod_{i\in U} e^{\ii (p_{+,i}-p_{-,i})\ell_i}\Big)
	\\
&	=
	\sum_n  {Q}_n(p_+) \sum_{0\le k  \le n }C_n^k \Big((\ii\partial_{p_+})^{k} \braket{p_+|p_-}\Big)
	\prod_{i\in U}(-\ell_i)^{n_i-k_i} e^{\ii (p_{+,i}-p_{-,i})\ell_i}
	\\
&	=
	\sum_n  {Q}_n(p_+)\prod_{i\in U} e^{\ii (p_{+,i}-p_{-,i})\ell_i} \(\ii\partial_{p_{+,i}}-\ell_i\)^{n_i} \prod_{j\in U'}  \(\ii\partial_{p_{+,j}}\)^{n_j}\braket{p_+|p_-}.
\end{split}
\ee
Here the set of all indices is represented as a disjoint union of two subsets, $\{1,2,...,D\} = U \sqcup U'$. The above expression allows us to obtain:
\be
Q_{\cal W}(x,p)
	= \sum_U \int_{\cal M} \D{k} e^{2\ii  x k} \braket{p+k|  \hat{Q}_{-U}|p-k}
	\prod_{i\in U} e^{2\ii  k_{i}\ell_i},
\ee
where
$$
{\hat{Q}_{-U}}\equiv
	\cQ(\ii\partial_p-n_U,{p})
	=
	\sum_n  {Q}_n(p_+)\prod_{i\in U} \(\ii\partial_{p_{+,i}}-\ell_i\)^{n_i} \prod_{j\in U'}  \(\ii\partial_{p_{+,j}}\)^{n_j}
$$
and by $n_U$ we denote vector with {$n_i = \ell_i$} for $i\in U$  and $n_i = 0$ for $i\in U^\prime$. We come to the final result:
\be\begin{split}
Q_{\cal W}(x,p)
	&= \sum_U \int_{\cal M} \D{k} e^{2\ii  (x+n_U) k} \braket{p+k|  \hat{Q}_{-U}|p-k}
	\\
&	=
	\sum_U Q_{-U,\cB}(x+n_U,p),
	\label{QO}
\end{split}
\ee
where the sum is, once again, over all possible subsets of $\{1,...,D\}$, while $Q_\cB$ is defined in \Ref{Q-Buot} as
$$
Q_\cB (x,p) = \int_{\cal M} \D{k} e^{2\ii  x k} \braket{ p+k|  \hat{Q}|p-k}.
$$
It is discussed in Appendix \ref{app:buot}.

{In the similar way we can represent $Q_\cW$ as follows
\be
\begin{split}
Q_{\cal W}(x,p)
&	=
	\sum_U \int_{\cal M} \D{k} e^{2\ii  (x-n_U) k} \braket{ p+k|  \hat{Q}_{U}|p-k}\\
&	=
	\sum_U Q_{U,\cB}(x-n_U,p).
\label{QO2-a}
\end{split}
\ee}
One can see, that the given operator $\hat Q$ defines uniquely the values of $Q_{\cW}(x,p)$ for $x\in \fO$. However, above it was argued that the given operator does not define uniquely the function $Q_{\cW}(x,p)$ for any real-valued $x$. One can see, that this ambiguity is related to the ambiguity of the analytical continuation of $Q_{\cW}(x,p)$ from its values at $x\in \fO$ to the real-valued $x$.

\subsubsection{Star identity I: $A_\cW\star B_\cB = (AB)_\cB$}
Let us now consider the Moyal product of two different symbols of operators $\hat A$ and $\hat B$, namely
\be
	r =	A_{\cal W}(x,p)\star B_\CB(x,p),\qquad
	\star=e^{-\frac\ii2\vec\partial_x \cev\partial_p+\frac\ii2\cev\partial_x\vec\partial_p}.
\ee
We substitute to this expression the definition of $B_\CB$ of Eq. \Ref{Q-Buot} and obtain:
\be
\begin{split}
r
&	=\frac{1}{2^D}\int_{\fM} \D{K}
		A_\cW(x+{\ii}\vec{\partial}_{p}/{2},{p}-{\ii} \vec{\partial}_x/{2})   e^{\ii  K x} B({p}+K/2,{p}-K/2)
		\\
&	=\frac{1}{2^D}\int_{\fM} \D{K}
		e^{\ii  Kx} A_{\cal W}(-\ii \cev{\partial}_K+{\ii}\vec{\partial}_{p}/{2},{p}+{K}/{2})B({p}+K/2,{p}-K/2) .
\end{split}
\ee
Using \Ref{corrl2} for $\hat A\equiv \cA (\ii\partial_p,p)$ with $f(P,K)=\delta(P-p)e^{\ii  Kx} $ and $h(P,K)=B(P+K/2,P-K/2)$ we further have
\begin{eqnarray}
r &=&
	\frac{1}{2^D}\int_{\fM} \D K e^{\ii  Kx} \cA (i {\partial}_K+{\ii}{\partial}_{p}/2,{p}+K/{2})   B({p}+K/2,{p}-K/2)\nonumber\\
&= &
	\frac{1}{2^D}\int_{\fM} \D K e^{\ii  Kx} \braket{ p+K/2|\hat{A}\hat{B}|p-K/2}
	\nonumber\\
& = & (AB)_\CB.
\end{eqnarray}
Thus we come to identity valid for $x \in \fO$
\begin{equation}
A_\cW(x,p)\star B_\cB(x,p)
	= (AB)_\cB(x,p)\label{ABWL}.
\end{equation}
In a similar way the following identity can be proved as well:
\begin{equation}
A_\cB(x,p)\star B_\cW(x,p) = (AB)_\cB(x,p), \qquad {x \in \fO}.
\label{ABWR}
\end{equation}
Namely, let us calculate \begin{eqnarray}
r =	A_\cB (x,p)\star B_{\cW}(x,p).
\end{eqnarray}
We substitute to this expression the above definition of $B_\CB$, Eq. (\ref{Q-Buot}), and obtain:
\be
r
	= \frac{1}{2^D}\int_{\fM}\D{P}  B_{\cal W}(x-{\ii}\vec{\partial}_{p}/{2},{p}+ {\ii} \vec{\partial}_x/{2})
	e^{\ii  {P} x} A({p}+{P}/2,{p}-{P}/2),
\ee
where $A(p,q) = \braket{ p|\hat{A}|q}$. We denote now $A^T(p,q) = A(q,p)= \braket{ p|\hat{A}^T|q}$, this gives
\be
r
	=\frac{1}{2^D}\int_{\fM} \D{P} e^{\ii  {P} x} B_{\cal W}(-\ii \cev{\partial}_{P}- {\ii} \vec{\partial}_{p}/{2},{p}- {P}/{2})A^T({p}-{P}/2,{p}+{P}/2).
\ee
The integration here is over $\fM$ and the integration by parts can be performed ({unlike Eq. (\ref{app:r1})}). Without loss of generality {we assume that ${\hat B}(i {\partial}_{p},{p})$ is written in the symmetric form: in each term the product of operators $\partial_p$ and $p$ is symmetrized.} {The corresponding representation can be obtained starting from Eqs. \Ref{Q1}, \Ref{Q2} applying the commutation relation between operators $\partial_p$ and $p$.} Then $\hat{B}^T(\ii {\partial}_{p},{p}) = \hat{B}(-\ii {\partial}_{p},{p})$. We come to
\begin{eqnarray}
r
	&=& \frac{1}{2^D}\int_{\fM} \D{P} e^{\ii  {P} x} \hat{B}(\ii {\partial}_{P}-{\ii}{\partial}_{p}/{2},{p}-{P}/{2})   A^T({p}-{P}/2,{p}+{P}/2)\nonumber\\
&=&
	\frac{1}{2^D}\int_{\fM} \D{P} e^{\ii  {P} x} \hat{B}^T(-\ii {\partial}_{P}+{\ii}{\partial}_{p}/{2},{p}-{P}/{2})   A^T({p}-{P}/2,{p}+{P}/2)\nonumber\\
&= &
	\frac{1}{2^D}\int_{\fM} \D{P} e^{\ii  {P} x} \braket{ p-P/2|\hat{B}^T\hat{A^T}|p+P/2}
\nonumber\\
& = &
	\frac{1}{2^D}\int_{\fM} \D{P} e^{\ii  {P}x} \braket{ p+P/2|\hat{A}\hat{B}|p-P/2} \nonumber\\
& = & (AB)_\CB.
\end{eqnarray}

\subsubsection{Star identity II: $A_{\cal W}\star B_{\cal W} = (AB)_{\cal W}$}

The above definition \Ref{Q_C-def} of the symbol $Q_{\cal W}$ of an operator $\hat Q$ through the series in powers of $x$ works not only for the lattice models, but for the continuous theories as well (it actually originated there). Then in Eq. (\ref{corrl2}) instead of $\fM$ one integrates over infinite momentum space $\dR^D$. Correspondingly, one again represents $Q_{\cal W}$ as
\be
{Q}_{\cal W}\(x,p\) = \sum_n x^n q_{n}(p)\label{Q2_}
\ee
with
\be
q_n(p) = \sum_{k \ge n} C_{k}^{n}\(\frac{\ii}{2}\partial_p\)^{k-n} Q_n(p)\label{Q3_}
\ee
Now $Q_n(p)$ depends on $p\in\dR^D$, and is not periodic. The derivation of Sect.  \ref{QWI} can be easily adopted to a continuum theory to give for any values of $x$ the following representation of the symbol of an operator:
\be
Q_{\cW}(x,p)
	= \int_{\dR^D} \D k e^{\ii  x k} \braket{ p+k/2|  \hat{Q}|p-k/2}.
\label{AWBC}
\ee
Notice, that unlike the lattice case in the continuum theory the representation through the series defines the Weyl symbol uniquely. The above representation allows to prove the following identity (in the way similar to the corresponding proof given in Appendix \ref{app:buot} for the Buot's symbol)
\be
A_{\cal W}(x,p)\star B_{\cal W}(x,p)
	= (AB)_{\cal W}(x,p). \label{ABstar}
\ee
This identity in continuum theory is valid for any continuous values of $x$. Looking at Eq. (\ref{ABstar}) one recognizes that when the symbol of operators is written as a series, Eq. \ref{ABstar} has purely algebraic nature, i.e. it contains the series in the derivatives (entering the star operation) and the series in powers of $x$ for both $A_\cW$ and $B_\cW$. Therefore, Eq. (\ref{ABstar}) should be valid for both continuous and lattice systems.

Let us now give a direct  proof of this identity in one-dimensional case. We have from \Ref{QO}
\be
\begin{split}
A_{\cal W}(x,p)
	&=  \sum_{n=0,1} A_{-n ,\cB}(x+n\ell,p),\\
B_{\cal W}(x,p)
	&=  \sum_{n=0,1} B_{-n,\cB}(x+n\ell,p),
\label{QO2}
\end{split}
\ee
where $\hat{A}_{-1 }(\ii\partial_p,p) = {\cal A}(\ii\partial_p-\ell,p)$ and $\hat{B}_{-1 }(\ii\partial_p,p) = {\cal B}(\ii\partial_p-\ell,p)$. Notice, that according to Eq. (\ref{starlattice}) in order to define the star product $A_\CB(x,p)\star B_\CB(x,p)$ at $x \in \fO$  we need to know the values of the functions $A_\CB(x,p)$ and $B_\CB(x,p)$ on $x \in \fO$ only. Therefore,
\be
(AB)_\CB(x,p)
	= A_{\cal W}(x,p) \star B_\cB (x,p) =  A_\cB (x,p) \star B_\cB (x,p) + A_{-1,\cB}(x+\ell,p) \star B_\cB (x,p).
\ee
At the same time in Appendix \ref{app:buot}, Eq. (\ref{Z2}) it is shown that $A_\CB\star B_\CB = (AB)_\CB$. Therefore, we conclude, that  $A_{-1,\cB}(x+\ell ,p) \star B_\cB (x,p)=0$ at $x \in \fO$ for arbitrary operators $\hat A$ and $\hat B$. In a similar way we obtain $A_\cB (x,p) \star B_{-1,\cB}(x+\ell ,p)=0$. Then,
\be
\begin{split}
A_{\cal W}(x,p)\star B_{\cal W}(x,p)
	& =  A_\cB (x,p) \star B_\cB (x,p) + A_{-1,\cB}(x+\ell ,p) \star B_{-1,\cB}(x+\ell ,p)\\
	& = (AB)_\cB (x,p) + (AB)_{-1,\cB}(x+\ell ,p)  = (AB)_{\cal W}.
 \label{(AB)_cW}
\end{split}
\ee
The condition, which we encountered this way, $A_{-1,\cB}(x+\ell ,p) \star B_\cB (x,p)=0$, is deeply non-trivial. We note that a more general form must hold true
\be
	A_{\cB}(x+\ell ,p) \star B_\cB (x,p)=0.
	\label{A-1*B}
\ee
Indeed, shifting the first  argument of ${\cal A}(\ii\partial_p,p)$ by a constant can only reshuffle the terms of the series expansion \Ref{Q1}, thus producing just another arbitrary operator. The equation in the form \Ref{A-1*B} is proved in \Ref{A-1*B-proof}.

Thus we have proven Eq. (\ref{ABstar}) in one dimensional case. The proof in the multidimensional case is completely similar. The validity of Eq. (\ref{ABstar}) can also be checked using the representation of Eqs. (\ref{Q1}), (\ref{Q2}), (\ref{Q3}). Upon substitution of those expressions to both sides of Eq. (\ref{ABstar}), it can be checked using Mathematica package that both sides are equal to each other when the finite number of terms in the powers of $x$ are taken.

\subsubsection{Trace definition}
Let us study the trace of an operator, given by its series \Ref{Q1}
\begin{eqnarray}
\tr \hat Q
	\equiv \int_\cM \D{p} \braket{p|\hat{Q}|p}
& = &
	\sum_n\int_\cM \D{p} \braket{p|(\ii\partial_p)^n Q_n(p)|p}   =
	\sum_{x\in\cO} \sum_n\int_\cM \D{p} \braket{ p | (\ii\partial_p)^n |x } \braket{ x |p } Q_n(p) \nonumber\\
& = &\frac{1}{|{\cM}|}
	\sum_{x\in\cO} \sum_n\int_\cM \D{p}  x^n Q_n(p)
 = \frac{1}{|{\cM}|}
	\sum_{x\in\cO} \int_\cM \D{p} Q_{\cal W}(x,p).
\label{TR}
\end{eqnarray}
We consider here one-dimensional case for simplicity. The trace over indices corresponding to any inner symmetries is implied, as before.
The cross check in coordinate space gives, using \Ref{QO2} and \Ref{GWxAH}
\be
\begin{split}
\frac{1}{|{\cM}|}
\sum_{x\in \cO} &\int_\cM \D{p} Q_{\cal W}(x,p)
 = \frac{1}{|{\cM}|}
	\sum_{x,z_1,z_2\in \cO} \int_\cM \D{p} \braket{ z_1| \hat{Q} |z_2} \delta_{2x,z_1+z_2} e^{\ii p (z_1-z_2)} \\&+\frac{1}{|{\cM}|}
	\sum_{x,z_1,z_2\in \cO} \int_\cM \D{p} \braket{ z_1| \hat{Q}_{-1} |z_2} \delta_{2(x+\ell),z_1+z_2} e^{\ii p (z_1-z_2)}\\
& =
	\sum_{x,z_1,z_2\in \cO}  \braket{ z_1| \hat{Q} |z_2} \delta_{2x,z_1+z_2} e^{\ii p (z_1-z_2)}\delta_{z_1,z_2} +\sum_{x,z_1,z_2\in \cO}  \braket{ z_1| \hat{Q}_{-1} |z_2} \delta_{2(x+\ell),z_1+z_2} e^{\ii p (z_1-z_2)}\delta_{z_1,z_2}\\
& =
	\sum_{x\in \cO}  \braket{ x| \hat{Q} |x}
	= \tr \hat Q.
\end{split}
\ee
 We might define the trace operation as
\be
\Tr_\cO Q_\cW
	\equiv \frac{1}{|{\cM}|}\sum_{x\in \cO} \int_\cM \D{p} Q_{\cal W}(x,p).
\label{cW-Tr-cO}
\ee
However, it does not satisfy \Ref{Tr-def-2}, as we will show in the next subsection.

An alternative definition, which does respect  \Ref{Tr-def-2}, is
\be
\Tr_\fO Q_\cW
\equiv \frac{1}{|{\fM}|}\sum_{x\in \fO} \int_\cM \D{p} Q_{\cal W}(x,p)
= \frac12\( \Tr_\cO Q_\cW (x,p)
+\Tr_{\cO'} Q_\cW(x,p)\),
\label{Tr_fO_cW}
\ee
where we recalled that in 1D we have $\fO=\cO\cup\cO'$, $\cO'=\cO+\ell$ (see Eq. \ref{fO}). Then
\be
\begin{split}
2\Tr_\fO Q_\cW(x,p)
	& = \Tr_\cO Q_\cW (x,p)
	+\Tr_{\cO'} Q_\cW(x,p) \\
&	=
	\tr \hQ
	+\Tr_{\cO} Q_\cW(x+\ell,p).
\end{split}	
\ee
Further applying \Ref{QO} (or alternatively, Eq. \ref{QO2}) we come to
\be
\begin{split}
2\Tr_\fO Q_\cW(x,p)
	& =
	\tr \hQ
	+\Tr_{\cO}Q_\cB(x+\ell,p)+\Tr_{\cO}Q_{-1,\cB}(x+2\ell,p)\\
&	=
	\tr \hQ
	+\tr\hQ_{-1},
	\label{cW-Tr-Q-Q1}
\end{split}	
\ee
where we used \Ref{cB-Tr-def-1} and \Ref{Tr_cO_ell}.

Thus, for the class of operators satisfying
\be
	\hQ =\hQ_{-1}
\label{op-class0}
\ee
the Weyl trace is given by
\be
\Tr_\fO Q_\cW(x,p)
	=\frac{1}{2|{\cM}|}\sum_{x\in \fO} \int_\cM \D{p} Q_{\cal W}(x,p).
	\label{cW-Tr-def0}
\ee
Again, the consideration in multi-dimensional case is similar, and for arbitrary $D$ we obtain
\be
\Tr_\fO Q_\cW(x,p)
	=\frac{1}{2^D|{\cM}|}\sum_{x\in \fO} \int_\cM \D{p} Q_{\cal W}(x,p).
	\label{cW-Tr-def}
\ee
for operators satisfying
\be
\hQ =\hQ_{-U}\qquad \forall U\subset \{1,2\ldots,D\}.
\label{op-class}
\ee
Note that $2^D|{\cM}| = |\fM|$.

\subsubsection{Trace of a product}

Finally, we will check Eq. \Ref{Tr-def-2}, i.e. if the star between two symbols of operators can be removed from an expression standing inside the trace. In the one-dimensional case (for simplicity) we can use \Ref{QO2} for each operator's symbol of a product to write
\be
\begin{split}
\Tr_\fO  \(A_\cW B_\cW\)
&	= \Tr_\fO \[A_\cB (x,p) B_\cB (x,p)\] +
	\Tr_\fO \[A_{-1,\cB}(x+\ell,p) B_{-1,\cB}(x+\ell,p)\],
\end{split}
\ee
where we used one of the remarkable properties of the $\cB$-symbol, namely \Ref{cB-Tr-Aell-B}, to get rid of the mixed terms of the type of $A_\cB(x,p) B_\cB(x+\ell,p)$. Applying next \Ref{cB-Tr-def-2} for each of the terms above, along with \Ref{Tr x+l} for the second one, we obtain
\be
\begin{split}
	2\Tr_\fO  \(A_\cW B_\cW\)
	= \tr \hat{A}\hat{B} + \tr \hat{A}_{-1}\hat{B}_{-1}.
\end{split}
\ee
On the other hand, using \Ref{cW-Tr-Q-Q1} for $\hat Q=\hat A \hat B$ (and thus $Q_\cW=A_\cW\star B_\cB$ due to Eq. \ref{(AB)_cW}) we immediately establish
\be
2\Tr_\fO  \(A_\cW \star B_\cW\)
	=2\Tr_\fO  \(A_\cW B_\cW\)
	=\tr \hat A \hat B+
	\tr \hat A_{-1} \hat B_{-1}.
\ee
We come to the identity
\be
\Tr_\fO  \(A_\cW \star B_\cW\)
	=\Tr_\fO  \(A_\cW B_\cW\).
\ee
Its validity in the multi-dimensional case can be proved in similar way. At the same time if $\hat{A}_{-U} = \hat{A}$ and $\hat{B}_{-U} = \hat{B}$ for any $U\subset \{1,2,\ldots D\}$, then
\be
\Tr_\fO  \(A_\cW \star B_\cW\)
	=\Tr_\fO  \(A_\cW B_\cW\) = {\rm tr}\, \hat{A}\hat{B}.
\ee
This compleets the proof of Theorem \ref{th-Q-series}.

\subsection{Wigner-Weyl formalism on a doubled lattice. One-dimensional case.}

\subsubsection{Operators on auxiliary lattice $\fO$ and their Weyl symbols}

In  Eq.  \Ref{fO} we introduced the extended auxiliary lattice $\fO$ that contains our physical lattice $\cO$ as a subset,
$$
\fO=\{ \ell k, k\in \dZ\} = \cO\cup\cO',
$$
$$
\cO= \{2\ell k, k\in \dZ\}, \quad \cO'=\{2\ell (k+1/2), k\in \dZ\}.
$$
To build a lattice Wigner transformation we now enlarge operators acting on $\cO$ to act on $\fO$. To this end we first introduce additional position and momenta eigenstates,
\be
\hat 1_\fO=\sum_{x\in\fO} \ket{x}\bra{x}= \int_{\fM}\D{p} \ket{p}\bra{p},\quad
\braket{x|p}=\frac1{\sqrt{|\fM|}} e^{\ii x p}, \quad \braket{x|y}= \delta_{x,y},
\ee
and continue physical operators to the auxiliary lattice by the following relations,
\be
\braket{x|\hQ|y}=\braket{x+\ell|\hQ|y+\ell},\quad x,y\in\cO.
\label{cond-2}
\ee
Moreover, we demand that the inter-lattice matrix elements vanish
\be
\braket{x|\hQ|y}=\braket{y|\hQ|x} = 0\quad \forall x\in\cO,y\in\cO'.
\label{cond-1}
\ee
The extended Fourier decomposition reads
\be
\ket{p}  = \frac1{\sqrt{|\fM|}}\sum_{x \in \fO}e^{\ii p x}\ket{x}
=   \frac1{\sqrt{{2\pi /\ell}}} \sum_{x\in \cO}e^{\ii p x}\ket{x}
+  \frac1{\sqrt{{2\pi/ \ell}}} \sum_{x\in \cO' }e^{\ii p x'}\ket{x'}.
\label{ket p}
\ee


Then, the matrix elements in momentum space of such operators become
\be
\begin{split}
\braket{p| \hQ |q}
& =
	\frac{1}{2\pi/\ell}\sum_{x_1,x_2\in \cO}\bra{x_1} \hat Q \ket{x_2}
		e^{\ii(x_2 q -x_1 p) }
	+ \frac{1}{2\pi/\ell}\sum_{x'_1,x'_2\in \cO'}\bra{x'_1} \hat Q \ket{x'_2}
		e^{\ii(x'_2 q-x'_1 p)} \\
& =
	\frac{1}{2\pi/\ell}\sum_{x_1,x_2\in \cO}\bra{x_1} \hat Q \ket{x_2}
	e^{\ii(x_2 q -x_1 p) }
	+ \frac{1}{2\pi/\ell}\sum_{x_1,x_2\in \cO}\bra{x_1+\ell} \hat Q \ket{x_2+\ell}
	e^{\ii(x_2 q-x_1 p)}e^{\ii(q- p) \ell } \\
& = \frac12 Q(p,q) (1+e^{\ii(q- p) \ell }),
\label{Q(p,q)}
\end{split}
\ee
where by definition
\be
  Q(p,q) = \frac{1}{|\cM|}\sum_{x_1,x_2\in \cO}\bra{x_1} \hat Q \ket{x_2}
	e^{\ii(x_2 q -x_1 p) }.
	\label{Q-Fourier}
\ee
Here $|\cM|=|\fM|/2=\pi/\ell$ (for the lattice in $1$D case). This definition, in particular, implies that $  Q(p,q)$ is periodic with the period $\pi/\ell$ in each of the arguments, since $\cO$ is a lattice with the link length being equal to $2\ell$, but the original matrix element, $\bra{p} \hat Q \ket{q} $, is still periodic with the period $2\pi/\ell$ due to the additional exponential factor in \Ref{Q(p,q)}.

Thus, the trace of such operator becomes
\be
\tr_\fO\hQ\equiv \sum_{x\in\fO} \braket{x|\hQ|y}
	=  2 \tr \hQ
	\label{tr-x}
\ee
On the other hand,
\be
\tr_\fM \hQ \equiv
	\int_\fM \D q \braket{q| \hQ |q}
	= \int_\fM \D q   Q(q,q)
	= 2 \int_\cM \D q   Q(q,q)
	=
	\int_{\cM} \D{p} \frac{1}{|\cM|}\sum_{x_1,x_2\in \cO} \,\bra{x_1} \hat Q \ket{x_2}
	e^{\ii(x_2  -x_1) p }
	=  2 \tr \hQ.
\label{tr-p}
\ee
Here by $\tr$ we denoted the ordinary, physical, trace of an operator.

\subsubsection{$W$-symbol and its properties}

We now use the $\cB$-symbol of operators, but defined on the extended lattice $\fO$. Using Eq. \Ref{Q(p,q)} we can express it as follows
\be
Q_W(x,p) \equiv
\int_{\fM} \D{q} e^{2 \ii q x} \braket{p+q|\hQ| p-q}
=
\frac12\int_{\fM} \D{q} e^{2 \ii q x}
  Q(p+q,p-q) (1+e^{-2 \ii q\ell}).
\label{Q-Weyl0}
\ee

Formally speaking, $Q_W(x,p)$ is defined for any $x\in\dR$.  However, for the discrete values, $x\in \fO$ we get a simpler expression. Indeed, due to the structure of the argument of the exponential factor in \Ref{Q(p,q)}, the integrand of \Ref{Q-Weyl0} is periodic with the period $\pi/\ell$, and thus the integration can be reduced from $\fM$ to $\cM$,
\be
Q_W(x,p) 		
	 = \int_{\cM} \D{q} e^{2 \ii q x}
  Q(p+q,p-q) (1+e^{-2 \ii q\ell})
	\label{Q-Weyl}
\ee
This is, therefore, the new definition of the Weyl symbol of an operator $\hQ$ defined on the lattice $\cO$. The second sublattice $\cO^\prime$ is an auxiliary instrument that may actually be omitted in the following as well as the extended lattice $\fO$.

Further, using Eq. \Ref{QO2-a} in \Ref{Q-Weyl}, we obtain
\be
\begin{split}
Q_W(x,p) &
	= Q_\CB(x,p) + Q_\CB(x-\ell,p)
	\\
&	=
	Q_\CB(x,p) + Q_{1,\cB}(x-\ell,p) + Q_\CB(x-\ell,p)-Q_{1,\cB}(x-\ell,p)\\
&	=
	Q_{\cal W}(x,p) + Q_\CB(x-\ell,p)-Q_{1,\cB}(x-\ell,p)
\end{split}
\ee
One can see, that $Q_W = Q_\cW$ for the operators of the class \Ref{op-class} considered in the previous Section, with $\hat{Q}_{1} = \hat{Q}$.

The inverse transformation reads
\begin{eqnarray}
\frac{1}{|\cM |}
\sum_{x\in \fO}e^{-\ii k x}Q_W(x,p) 		& = &  \frac{1}{|{\cal M}|}  \sum_{x\in \fO}\int_{\cM} \D{q} e^{2 \ii (q-k/2) x}
	  Q(p+q,p-q) (1+e^{-2 \ii q\ell})\nonumber\\
&=&
	\frac{1}{|{\cal M}|}\frac{\pi}{\ell} \int_{\cM} \D{q} \delta^{[\pi/\ell]}
	  Q(p+q,p-q) (1+e^{-2 \ii q\ell})\nonumber\\
&=&
     Q(p+k/2,p-k/2)\(1+e^{-2 \ii\(k/2 \bmod{\pi/\ell}\)\ell}\),
   \label{iQ}
\end{eqnarray}
{where by definition
	$\delta^{[a]}(q) = \sum_{N\in \dZ} \delta(q - a N) $.} It gives
\begin{eqnarray}
  Q(P_1,P_2) &=&
\frac{1}{|\cM|\Big(1+e^{-2 \ii ( P_1-P_2 )\ell} \Big)}\sum_{x\in \fO}e^{-\ii (P_1-P_2) x}Q_W(x,(P_1+P_2)/2).
\label{iQ_}
\end{eqnarray}

It is worth mentioning, that the definition of the Weyl symbol of operator $\hat Q$ given by Eq. (\ref{Q-Weyl}) obeys the following important property:
$$
Q_W(x + \ell,p) = \Big(\hat{Q}^T\Big)_W(-x,p).
$$

\subsubsection{Moyal product}

We know that the $\cB$-symbol does map the product of operators into the star product of their symbols. Thus, the same property is naturally expected from Eq. \Ref{Q-Weyl} based on the definition of the $\cB$-symbol,
\be
	(\hat A \hat B)_W = A_W\star B_W.
	\label{tilde-star}
\ee
Written in terms of the Fourier transform of Eq. \Ref{Q-Fourier}, it becomes
\be
\int_{\cM}\D{q} e^{ 2 \ii  q x} (\hat A \hat B)(p+q,p - q)  f(q)
	=
	\(\int_{\cM}\D{q}e^{ 2 \ii  q x}
	  A(p+q,p - q)
	f(q)\)\star
	\(\int_{\cM}\D{q'} e^{ 2 \ii  q' x}
	  B(p+q',p - q')
	f(q')\),
\ee
where we denoted $f(q)= 1+e^{-2 \ii q\ell}$ for brevity. Both $f(q)$ and $  A$, $  B$ are periodic with the period  $\pi/\ell$, while  $\cM=(-\pi/2\ell,\pi/2\ell]$. It is instructive to prove the above equation without referring to the extended lattice $\fO$, i.e. working within $\cO$ only.

The RHS of Eq. (\ref{tilde-star}) reads
\be
\begin{split}
A_W\star B_W  &=
	\int_{-\pi/2\ell}^{\pi/2\ell}\D{q d q'}
	f(q)f(q')
	  A(p+q,p - q)
	e^{ 2 \ii  q x}
	e^{-\frac\ii2\vec\partial_x \cev\partial_p+\frac\ii2\cev\partial_x\vec\partial_p}
	e^{2 \ii  q' x}
	  B(p+q',p - q') \\
	&=
	\int_{-\pi/2\ell}^{\pi/2\ell}\D{q d q'}e^{ 2 \ii  q x}e^{2 \ii  q' x}
	f(q)f(q')
	  A(p+q,p - q)
	e^{q'\cev\partial_p-q\vec\partial_p}	
	  B(p+q',p - q')  \\
	&=
	\int_{-\pi/2\ell}^{\pi/2\ell}\D{q d q'}e^{ 2 \ii  q x}e^{2 \ii  q' x}
	f(q)f(q')
	  A(p+q+q',p - q+q')
	  B(p-q+q',p -q- q')
\end{split}	
\ee
\begin{figure}
    \centering
	\begin{minipage}{0.45\textwidth}
		\centering
		\includegraphics{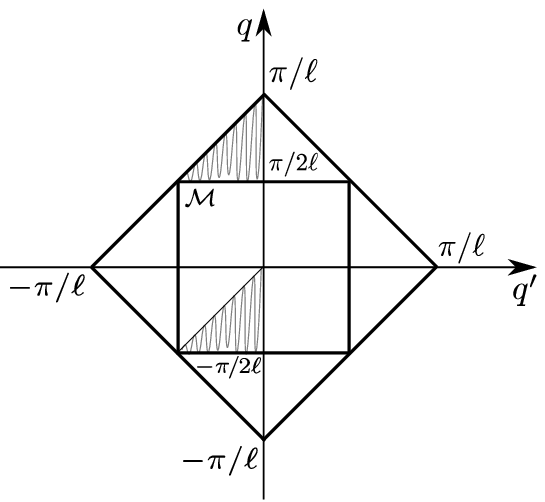}
		\caption{Transformation of an integral over square to a rhombus: shaded regions give equal contributions.}
		\label{fig:rhomb}
	\end{minipage}\hfill
	\begin{minipage}{0.45\textwidth}
		\includegraphics{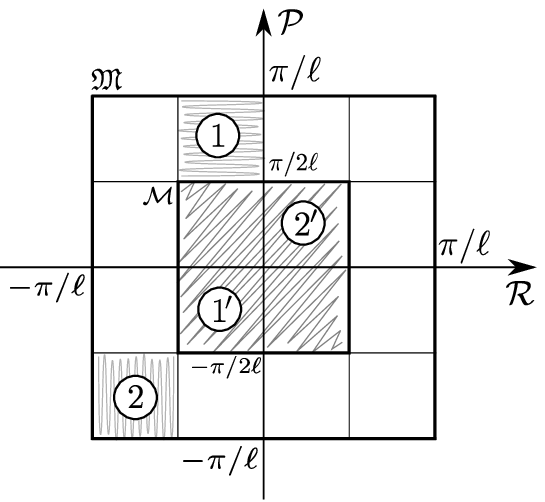}
		\caption{Double, $\fM\times\fM$, and simple, $\cM\times\cM$, Brillouin zones.}
		\label{fig:square}
	\end{minipage}
\end{figure}
Transforming the integration area to a rhombus, see Fig. \ref{fig:rhomb}, and changing the variables
\be
	\cP=q+q',\quad \cR = q-q'
\ee
we have
\be
\begin{split}
	A_W\star B_W &=
	\frac1{2}	\iint_{\diamondsuit}\D{q d q'}e^{ 2 \ii  \cP x}
	f(q)f(q')
	  A(p+q+q',p -q+q')
	  B(p-q+q',p -q-q') \\
	&=
	\frac{1}{4}	\int_{-\pi/\ell}^{\pi/\ell} \D{\cP d \cR}e^{ 2 \ii  \cP x}
	f(\tfrac{\cP+\cR}2)f(\tfrac{\cP-\cR}2)
	  A(p+\cP,p - \cR)
	  B(p-\cR,p -\cP).
\end{split}	
\ee
We used that the Jacobian of the transformation is  $|J|=1/2$. Notice that  we cannot reduce back the integration area from $\fM=({-\pi/\ell},{\pi/\ell}]$ to $\cM=({-\pi/(2\ell)},{\pi/(2\ell)}]$ because  $f(\tfrac{\cP\pm \cR}{2})$ is periodic with the period $2\pi/\ell$ as a function of $\cP$ or $\cR$, while $  A$, $  B$ are periodic with the period $\pi/\ell$. We introduce now
\be
g(p)\equiv f(p)-1 = e^{-2 \ii q\ell},
\ee
then
\be
\begin{split}
	A_W\star B_W &=
	\frac{1}{4}	\int_{-\pi/\ell}^{\pi/\ell} \D{\cP d \cR}e^{ 2 \ii  \cP x}
	\[1+g(\tfrac{\cP+\cR}2)+g(\tfrac{\cP-\cR}2)+g(\tfrac{\cP+\cR}2)g(\tfrac{\cP-\cR}2)\]
	  A(p+\cP,p - \cR)   B(p-\cR,p -\cP).
\end{split}	
\ee
Noting that $g(p\pm\pi/(2\ell))= -g(p)$, we see that the first and the last terms in the square parenthesis do not change their signs when either $\cP$ or $\cR$ is shifted by $\pi/\ell$. Thus, these terms give rise to  $4$ times the integral over $\cM= ({-\pi/(2\ell)},{\pi/(2\ell)} ]$.

The behavior of the remaining terms in the square parenthesis is more complicated. The shift of squares of type $1$ on Fig. \ref{fig:square} by $\pi/\ell$ in the corresponding argument (i.e. to the position denoted by $1'$)  changes the sign of the $g$-factor, while shifting of the squares of type $2$ into position $2'$ does not. Consequently, the net result of the integration of these two terms gives zero -- eight positives squares (4 diagonal and 4 inner ones), and eight negatives ones (the outer off diagonal squares).
Thus, we come to
\be
\begin{split}
A_W\star B_W &=
	\int_{-\pi/2\ell}^{\pi/2\ell} \D{\cP d \cR}e^{ 2 \ii  \cP x}
	\[1+g(\tfrac{\cP+\cR}2)g(\tfrac{\cP-\cR}2)\]
	  A(p+\cP,p - \cR)   B(p-\cR,p -\cP).
	\label{A*B-1}
\end{split}	
\ee
In 1D we trivially have
\be
1+g(\tfrac{\cP+\cR}2)g(\tfrac{\cP-\cR}2) =
1+g(\cP) = f(\cP),
\ee
since $g(q)=e^{-2 \ii q\ell}$. To finalize the procedure we study
\be
\begin{split}
	\int_{\cM} \D{k}   A(p,k)   B(k,q)
	& 	= \frac{1}{(\pi/\ell)^2} \int_{\cM} \D{k} \sum_{x_1,x_2\in \cO\atop y_1,y_2\in \cO}
	\bra{x_1} \hat A \ket{x_2}
	e^{\ii(x_2 k -x_1 p) }	
	\bra{y_1} \hat B \ket{y_2} e^{\ii(y_2q -y_1k) }
	\\
	&	 = \frac{1}{\pi/\ell}  \sum_{x_1,x_2\in \cO\atop y_1,y_2\in \cO}
	\braket{x_1 | \hat A | x_2}
	e^{-\ii x_1p }	
	\bra{y_1} \hat B \ket{y_2} e^{\ii y_2 q } \delta(x_2-y_1)\\
	&	 = \frac{1}{\pi/\ell}  \sum_{x_1,y_2\in \cO}
	\braket{x_1 | \hat A \hat B | y_2} e^{\ii y_2 q-\ii x_1 p }	\\
	&	 = \widetilde{ A B } (p, q).
\end{split}
\label{ABdk}
\ee
This finishes the proof of \Ref{tilde-star}.

\subsubsection{Trace and its properties}

As before, we can introduce two possible trace operations. One is using the summation over the points of lattice $\cO$ only,
\be
\begin{split}
	\Tr_\cO Q_W
	&\equiv
	\frac{1}{|\cM|}\int_{\cM} \D{p}\sum_{x\in \cO}  Q_W(x,p) 		
	= \frac{1}{|{\cal M}|} \int_{\cM} \D{p} \sum_{x\in \cO}\int_{\cM} \D{q} e^{2 \ii q x}
	\,   Q(p+q,p-q) (1+e^{-2 \ii q\ell})\\
	&	=
	\int_{\cM} \D{p dq} \delta^{[\pi/\ell]}(2q)
	  Q(p+q,p-q) (1+e^{-2 \ii q\ell})\\
	&	=
	\int_{\cM} \D{p}
	\,   Q(p,p) = \tr\hat{Q}.
	\label{TrQ-cO}
\end{split}
\ee

However, one can also write
\be
\Tr_\fO Q_W = \frac{1}{|\fM|}\int_{\cM} \D{p}\sum_{x\in \fO} Q_W(x,p),
	\label{TrQ-fO}
\ee
and it is a simple exercise based on \Ref{TrQ-cO} to show that
\be
\Tr_\fO Q_W  =  \tr \hQ.
\ee

For the product of two operators $\hat{A}$ and $\hat{B}$ we have
\be
\begin{split}
	\Tr_\fO  &A_W(x,p) B_W(x,p) 	\\	
	&	
	=   \frac{1}{|\fM|} \int_{\cM} \D{p} \sum_{x\in \fO}\int_{\cM} \D{q} e^{2 \ii q x}
	\,   A(p+q,p-q) (1+e^{-2 \ii q\ell})
	\int_{\cM} \D{k} e^{2 \ii k x}
	\,   B(p+k,p-k) (1+e^{-2 \ii k\ell})
	\\
	& =\frac12   \int_{\cM} \D{pdq}
	   A(p+q,p-q) (1+e^{-2 \ii q\ell})
	  B(p-q,p+q) (1+e^{2 \ii q\ell})
	\\
	&	= \frac12\int_{\cM} \D{pdq}
 	  A(p+q,p-q)   B(p-q,p+q) (2+e^{2 \ii q\ell} + e^{-2 \ii q\ell} )\\
	 & =  \frac{1}{2|\cM|^2}\int_{\cM} \D{pdq}
	\sum_{x_{1,2},y_{1,2}\in \cO}\braket{x_1 | \hat A |x_2}
	e^{\ii(x_2 (p+q) -x_1 (p-q)) }  \braket{y_1| \hat B |y_2}
	e^{\ii(y_2 (p-q) -y_1 (p+q)) }(2+e^{2 \ii q\ell} + e^{-2 \ii q\ell} )
	\\
	&	= \frac12
	\sum_{x_1,x_2\in \cO}\braket{x_1| \hat A |x_2}
	\sum_{y_1,y_2\in \cO}\braket{y_1| \hat B |y_2}
	(2\delta_{2x_1,2y_2}\delta_{2x_2,2y_1}+\delta_{2x_1,2y_2+2\ell}\delta_{2x_2,2y_1+2\ell} + \delta_{2x_1,2y_2-2\ell}\delta_{2x_2,2y_1-2\ell} )
\end{split}
\ee
In the last expression the second and the third terms proportional to $ \delta_{2x_1,2y_2\pm2\ell}\delta_{2x_2,2y_1\pm 2\ell}$ vanish for $x_i, y_i \in \cO$, and we arrive at
\be
\Tr_\fO  A_W(x,p) B_W(x,p)
	=
	\sum_{x_1,x_2\in \cO}\braket{x_1| \hat A |x_2}
\braket{x_2| \hat B |x_1} = \tr \hat{A} \hat{B},
\ee
which together with \Ref{tilde-star} proves that $\Tr_\fO$ satisfies Eq. \Ref{Tr-def-2}, and to prove the whole of the {\it Theorem} \ref{th-Q-W} it only lacks demonstrate \Ref{id-deg}. It is immediate once we use \Ref{Q-Weyl0}.

\subsubsection{Groenewold equation}
Let us check now that introduction of the auxiliary lattice $\fO$ does not spoil anything.

We consider two operators $\hQ, \hG$ that obey Eqs. (\ref{cond-2}) and (\ref{cond-1}), which are inverse to each other:
\be
	\hQ \hG=\hat 1.
\ee
The Fourier representation gives
\be
	\int_{\fM} \D{q} \braket{p|Q|q} \braket{q|G|k} = \delta^{[2\pi/\ell]}(p-k) = \sum_{N\in\dZ} \delta(p-k +  (2\pi/\ell) N) ,
	\label{QG}
\ee
whose LHS using \Ref{Q-Fourier} can be rewritten as
\be
\begin{split}
\int_{\fM} \D{q} \braket{p|Q|q} \braket{q|G|k}  &=
	\frac14	\int_{\fM} \D{q}   Q(p,q)    G(q,k)  (1+e^{\ii(q- p) \ell })(1+e^{\ii(k- q) \ell }).
\end{split}	
\ee
To check the consistency, we calculate
\be
\begin{split}
\int_{\fM} \D{q}   Q(p,q)    G(q,k) &
	= \frac{1}{(\pi/\ell)^2}
	\int_{\fM} \D{q}\sum_{x_1,x_2\in \cO \atop y_1,y_2\in \cO}
		\bra{x_1} \hat Q \ket{x_2} 	e^{\ii(x_2 q -x_1 p) }
		\bra{y_1} \hat G \ket{y_2} 	e^{\ii(y_2 k -y_1 q) }\\
& 		= \frac{1}{(\pi/\ell)^2}
	\sum_{x_1,x_2\in \cO \atop y_1,y_2\in \cO}
	\bra{x_1} \hat Q \ket{x_2} 	\bra{y_1} \hat G \ket{y_2} 	e^{\ii(y_2 k -x_1 p) }
	\int_{\fM} \D{q} e^{\ii(x_2  -y_1 )q }
	\\
& 		= \frac{2\pi/\ell}{(\pi/\ell)^2}
	\sum_{x_1,x_2\in \cO \atop y_1,y_2\in \cO}
	\bra{x_1} \hat Q \ket{x_2} 	\bra{y_1} \hat G \ket{y_2} 	e^{\ii(y_2 k -x_1 p) }
	\delta(x_2  -y_1 )
	\\
& 		= \frac{2 }{\pi/\ell}
	\sum_{x_1,y_2\in \cO; x_2\in \fO}
	\bra{x_1} \hat Q \ket{x_2} 	\bra{x_2} \hat G \ket{y_2} 	e^{\ii(y_2 k -x_1 p) },
\end{split}	
\ee
in the last line we expanded the summation in $x_2$ to $\fO=\cO\cup\cO'$ using  \Ref{cond-2}. And now
\be
\begin{split}
\int_{\fM} \D{q}   Q(p,q)    G(q,k)
& 	= \frac{2 }{\pi/\ell}
	\sum_{x_1,y_2\in \cO}
	\bra{x_1} \hat Q \hat G \ket{y_2} 	e^{\ii(y_2 k -x_1 p) }
	= \frac{2 }{\pi/\ell}
	\sum_{x_1,y_2\in \cO}
	\delta_{x_1  y_2} 	e^{\ii(y_2 k -x_1 p) }
	\\	
& 	=  \frac{2 }{\pi/\ell}
	\sum_{x_1\in \cO}	e^{\ii(k -p)x_1}\\
&	
	= 2 \delta^{[\pi/\ell]}(k-p).
\end{split}	
\ee
Thus we have
\be
\begin{split}
\int_{\fM} \D{q} \braket{p|Q|q} \braket{q|G|k}
& 	=	\frac12
	\delta^{[\pi/\ell]}(k-p)\(1+e^{\ii(k-p)\ell }\)
	+
	\int_{\fM} \D{q}   Q(p,q)    G(q,k)  \(e^{\ii(k- q) \ell }+e^{\ii(q- p) \ell }\).
\end{split}	
\ee
We recall now that  ${\fM}=(-\pi/\ell,\pi/\ell]$ while $  Q(p,q)$ is periodic with the period $\pi/\ell$, and the exponential factor changes its sign under the shift by  $\pi/\ell$. Therefore, the last term in the above expression vanishes. We arrive at
\be
	\int_{\fM} \D{q}\braket{p|Q|q} \braket{q|G|k}=
	\frac12\(1 +  e^{\ii(k-p) \ell }\) \delta^{[\pi/\ell]}(k-p),
\ee
which is actually a representation of $\delta^{[2\pi/\ell]}(k-p)$ as in \Ref{QG}, while
\be
\int_{\cM} \D{q}   Q(p,q)   G(q,k)   = 	\delta^{[\pi/\ell]}(k-p).
\ee

Thus the Weyl transform given by \Ref{Q-Weyl} does  result in
\be
(\hQ \hG)_W = Q_W\star G_W =1
\ee

\subsection{Wigner-Weyl formalism on an extended lattice. The multidimensional case.}

\subsubsection{The auxiliary extended lattice}

Let us extend our consideration to the multidimensional case. We are going to obtain an  analogue of Eq. \Ref{Q-Weyl}. Rectangular lattice and its first Brillouin zone are
\be
	\fO=\left\{\sum_{j=1}^M k_j \belj,\ k_j\in\dZ\right\},\quad
	\fM = \left\{\sum_{j=1}^M \al_j \bgj,\ \al_j\in(-1/2,1/2]\right\},\qquad
	\belj\bg^{(i)}=2\pi \delta_{ij},\quad i,j=1,2,\ldots M.
\ee
$M=D-1$ is the number spatial dimensions.
Again, we separate the constituents of the original lattice
\be
\fO=\cO\cup \cO',\qquad
	\cO= \left\{\sum_{j=1}^M 2 k_j \belj, k_j\in\dZ\right\}, \quad
	\cO'=\left\{\sum_{j=1}^M k_j \belj, \exists i: k_i=2n+1;\ k_j,n\in\dZ\right\}.
\ee
It can also be written as $2^M$ copies of $\cO$:
\be
\fO = \bigcup_{ U}\cO_{ U},\quad
	\cO_{ U} = \left\{
		\sum_{j\in U} (2k_j+1) \belj
		+\sum_{j\not \in U} 2k_j\belj,\ k_j\in\dZ\right\}.
\ee
here $ U$ is any subset of $\{1,2,\ldots M\}$ including the empty one, the number of such subsets is $2^M$. The empty subset corresponds to $\cO$, evidently. The relation between the sublattices is now
\be
	\cO_{ U} =  \cO+\bel_{ U},
\ee
where
\be
	\bel_{ U} \equiv \sum_{j\in U}\bel^{(j)}.
\ee
and we assume that $\bel_\emptyset=\bm0$.

Along the lines of the $1$D-case, we consider $Q$-class of operators, which satisfies  the following condition
\be
	\braket{\bx|\hQ|\by}=\braket{\by|\hQ|\bx} = 0\quad \forall \bx\in\cO_{ U},\by\in\cO_{U'},
	 U\ne U',
	\label{cond-1-D}
\ee
which guarantees that the matrix elements are non-zero only for the transitions within  the same sublattice.

The second condition becomes
\be
\braket{\bx|\hQ|\by}
	=\braket{\bx+\bel_{ U} |\hQ|\by+\bel_{ U}} ,\qquad \forall \bx,\by\in\cO,\ \forall  U \subseteq\{1,2,\ldots M\}.
	\label{cond-2-D}
\ee

First, we note that in the $D$-dimensional case
all states $\ket{\bp}$ are periodic with the periods $\bgj$. The Fourier transform will read
\be
\begin{split}
\ket{\bp}  &
	= \frac1{\sqrt{|\fM|}}\sum_{\forall U} \sum_{\bx\in \cO_{ U} } e^{\ii \bp \bx}\ket{\bx}\\
	& =  \frac1{\sqrt{|\fM|}} \sum_{\bx\in \cO}e^{\ii \bp \bx}\ket{\bx}
	+  \frac1{\sqrt{|\fM|}} \sum_{  U\ne\emptyset} \sum_{\bx'\in \cO_{ U} }e^{\ii \bp \bx'}\ket{\bx'}.
\end{split}
\label{ket bp}
\ee
Using \Ref{cond-1-D} we obtain
\be
\begin{split}
	\bra{\bp} \hat Q \ket{\bq}
	& =
	 \frac1{|\fM|} \sum_{\bx_1,\bx_2\in \cO}\bra{\bx_1} \hat Q \ket{\bx_2}
	e^{\ii(\bx_2 \bq -\bx_1 \bp) }
	+\frac1{|\fM|}\sum_{  U\ne\emptyset} \sum_{\bx'_{1,2}\in \cO_{ U} } \bra{\bx'_1} \hat Q \ket{\bx'_2}
	e^{\ii(\bx'_2 \bq-\bx'_1 \bp)} \\
	& =
	\frac1{|\fM|}\sum_{\bx_1,\bx_2\in \cO}\bra{\bx_1} \hat Q \ket{\bx_2}
	e^{\ii(\bx_2 \bq -\bx_1 \bp) }
	+ \frac1{|\fM|}\sum_{  U\ne\emptyset} \sum_{\bx'_{1,2}\in \cO_{ U} }
	\braket{\bx_1+\bel_{ U} |\hQ|\bx_2+\bel_{ U}}
	e^{\ii(\bx_2 \bq-\bx_1 \bp)}e^{\ii(\bq- \bp) \bel_{ U}} \\
& 	= \frac1{2^D}  Q(\bp,\bq)
		\(1+\sum_{ U\ne\emptyset} e^{\ii(\bq- \bp) \bel_{ U} }\)
	\equiv 	\frac1{2^D}  Q(\bp,\bq)
		\sum_{\forall U} e^{\ii(\bq- \bp) \bel_{ U} }.
\end{split}
\ee
Here we introduced similarly to \Ref{Q-Fourier}
\be
  Q(\bp,\bq) \equiv  \frac{1}{|\cM|}
	\sum_{\bx_1,\bx_2\in \cO}\bra{\bx_1} \hat Q \ket{\bx_2}
	e^{\ii(\bx_2 \bq -\bx_1 \bp) }.
\ee
This object is periodic with the periods $\bgj/2$. We also represent
$$
\bra{\bp} \hat Q \ket{\bq} = \frac1{2^D}  Q(\bp,\bq) f(\tfrac{\bp- \bq}2),
$$
where
\be
f(\bq)\equiv\sum_{\forall U} e^{-2\ii\bq\bel_{ U} }
	= \prod_{j=1}^M \(1+e^{-2\ii\bq\belj }\).
	\label{f}
\ee

\subsubsection{Moyal product}
The multidimensional Weyl symbol shall be defined as
\be
Q_W(\bx,\bp) =   \frac1{2^D} \int_{\fM }\D{\bq} e^{ 2 \ii  \bq \bx}
	  Q(\bp+\bq,\bp - \bq)
	f(\bq)
	=   \int_{\cM}\D{\bq}e^{ 2 \ii  \bq \bx}
	  Q(\bp+\bq,\bp - \bq)
	f(\bq)
	\label{Q Weyl D}
\ee
for $\bx\in\fO$.  It satisfies all the properties we require for the proper Weyl symbol in Sect. \ref{SectAxiom}
Let us prove, as an example of  the $D$-dimensional considerations, that
\be
(\hat A\hat B)_W=A_W\star B_W.
\ee
In other words, we shall prove that
\be
\int_{\cM}\D{\bq} e^{ 2 \ii  \bq \bx}
(\hat A \hat B)(\bp+\bq,\bp - \bq)
f(\bq)
=
\(\int_{\cM}\D{\bq}e^{ 2 \ii  \bq \bx}
  A(\bp+\bq,\bp - \bq)
f(\bq)\)\star
\(\int_{\cM}\D{\bq'} e^{ 2 \ii  \bq' \bx}
  B(\bp+\bq',\bp - \bq')
f(\bq')\).
\ee
The RHS of this expression reads
\be
\begin{split}
RHS &=
	\int_{\cM}\D{\bq d \bq'}
	f(\bq)f(\bq')
	  A(\bp+\bq,\bp - \bq)
	e^{ 2 \ii  \bq \bx}
	e^{-\frac\ii2\vec\partial_\bx \cev\partial_\bp+\frac\ii2\cev\partial_\bx\vec\partial_\bp}	
	e^{2 \ii  \bq' \bx}
	  B(\bp+\bq',\bp - \bq') \\
&=
	\int_{\cM}\D{\bq d \bq'}e^{ 2 \ii  \bq \bx}e^{2 \ii  \bq' \bx}
	f(\bq)f(\bq')
	  A(\bp+\bq,\bp - \bq)
	e^{\bq'\cev\partial_\bp-\bq\vec\partial_\bp}	
	  B(\bp+\bq',\bp - \bq')  \\
&=
	\int_{\cM}\D{\bq d \bq'}e^{ 2 \ii  \bq \bx}e^{2 \ii  \bq' \bx}
	f(\bq)f(\bq')
	  A(\bp+\bq+\bq',\bp - \bq+\bq')
	  B(\bp-\bq+\bq',\bp -\bq- \bq').
\end{split}	
\ee
Then we introduce notation
\be
	\bcP=\bq+\bq',\quad \bcR = \bq-\bq'
\ee
and transform the integration area to the form of a multi-dimensional rhombus
\be
\begin{split}
A_W\star B_W &=
	\frac1{2^D}	\iint_{\diamondsuit}\D{\bq d \bq'}e^{ 2 \ii  \bcP \bx}
	f(\tfrac{\bcP+\bcR}2)f(\tfrac{\bcP-\bcR}2)
	  A(\bp+\bcP,\bp - \bcR)
	  B(\bp-\bcR,\bp -\bcP) \\
&=
	\frac{|J|}{2^D}	\int_{\fM}\D{\bcP d \bcR}e^{ 2 \ii  \bcP \bx}
	f(\tfrac{\bcP+\bcR}2)f(\tfrac{\bcP-\bcR}2)
	  A(\bp+\bcP,\bp - \bcR)
	  B(\bp-\bcR,\bp -\bcP).
	\label{diam-D}
\end{split}	
\ee
As in the 1D case we write
\be
	g(\bq)\equiv f(\bq)-1 = \sum_{U\ne\emptyset} e^{-2\ii\bq\bel_{U} }.
\ee
Then
\be
\begin{split}
	A_W\star B_W &=
	\frac{1}{4}	\int_{\fM} \D{\bcP d \bcR}e^{ 2 \ii  \bcP \bx}
	\[1+g(\tfrac{\bcP+\bcR}2)+g(\tfrac{\bcP-\bcR}2)+g(\tfrac{\bcP+\bcR}2)g(\tfrac{\bcP-\bcR}2)\]
	  A(\bp+\bcP,\bp - \bcR)
	  B(\bp-\bcR,\bp -\bcP)
\end{split}	
\ee
For each term of $g(\tfrac{\bcP\pm\bcR}2)$   there exists such $j$ that the shift $\bcP\to \bcP\pm\bgj/2$ (or similar with $\bcR$) changes its sign. Thus such a term will be cancelled when integrated in the $j$-th direction of $\fM$ (see Fig. \ref{fig:square}).

The product $g(\tfrac{\bcP+\bcR}2)g(\tfrac{\bcP-\bcR}2)$ requires an additional analysis. Indeed, it will contain the exponents of the type
\be
\exp\left\{-\ii\( \tfrac{\bcP+\bcR}2\bel_{ U} +\tfrac{\bcP-\bcR}2\bel_{U'} \)\right\}
	=\exp\left\{-\ii\(\bcP \bel_{ U\cap  U'}
			+\tfrac{\bcP+\bcR}2\bel_{ U\smallsetminus U'} +\tfrac{\bcP-\bcR}2\bel_{ U'\smallsetminus U} \)\right\},
\ee
now shifting $\bcP$ by $\bgj$ will affect it by the following factor
\be
\left\{\begin{array}{ll}
	e^{-\ii \bgj \belj}=1,\quad& j \in  U\cap  U' ,\\
	e^{-\ii \bgj \belj/2}=-1,& j \notin  U\cap  U' .
\end{array}
\right.
\ee
and similarly for the shifts of $\bcR$.  So, under $D$ shifts we transform $\fM$ to the smaller $\cM$, and only the  terms  with
\be
  U =  U'
\ee
survive.
Thus
\be
\begin{split}
A_W\star B_W &=
	\frac{4^D |J|}{2^D}	\int_{\cM}\D{\bcP d \bcR}e^{ 2 \ii  \bcP \bx}
	\(1+\sum_{ U\ne\emptyset} e^{-\ii\bcP\bel_{ U} } \)
	  A(\bp+\bcP,\bp - \bcR)
	  B(\bp-\bcR,\bp -\bcP) .
\end{split}	
\ee
The $\bcR$ integration goes exactly as in \Ref{ABdk}, and we arrive at
\be
(\hat A\hat B)_W
	= A_W\star B_W,
\ee
where each Weyl symbol is understood in terms of the operators acting in $\cO$ only, as in the RHS of \Ref{Q Weyl D}.

In the complete analogy with the one-dimensional case we can also  demonstrate all other relevant properties. In particular, for the operators that obey $\hat{A}_{-U} = \hat{A}$ we have
$A_W(x,p) = A_{\cal W}(x,p)$.

\subsection{A brief word on motivation}

To develop the double/extended  lattice approach we used as a motivation the following consideration.
Let us consider the following $\cB$-symbol in the extended lattice
\be
	Q_\cB =   \int_{\fM }d\bq\,\,e^{ 2 \ii  \bq \bx}
\braket{ \bp+\bq| \hat Q | \bp - \bq }
\ee
and apply it to an exponential operator
\be
\hQ = e^{\ii (\hat\bp \bb+\hat\bx \bl)},
\ee
where $\bb$ and $\bl$ are certain constant vectors, $\hat\bx\equiv- \ii\partial_\bp$. Then using the Hausdorff formula we obtain
\be
\[\hat\bp \bb, \hat\bx \bl\] = \ii \bb\bl \quad
\Rightarrow
\quad
e^{\ii (\hat\bp \bb+\hat\bx \bl)}
= e^{\ii \hat\bp \bb} e^{\ii\hat\bx \bl} e^{-\ii\bb\bl/2}
\ee
the $\cal B$-symbol becomes (since $e^{\ii\hat\bx \bl}= e^{+\bl \partial_\bp}$)
\be
\begin{split}
	Q_\cB &=  e^{-\ii\bb\bl/2} \int_{\fM }\D\bq e^{ 2 \ii  \bq \bx}
	\braket{ \bp+\bq| e^{\ii \hat\bp \bb} e^{\ii\hat\bx \bl}| \bp - \bq }  \\
	&	=  e^{-\ii\bb\bl/2} \int_{\fM } \D\bq e^{ 2 \ii  \bq \bx}
	\braket{ \bp+\bq| e^{\ii \hat\bp \bb} | \bp - \bq +\bl}  \\
	&	=  e^{-\ii\bb\bl/2} \int_{\fM } \D\bq e^{ 2 \ii  \bq \bx}
	e^{\ii ( \bp - \bq +\bl) \bb} \delta^{[\bg]}( \bp+\bq -(\bp - \bq +\bl)) \\
	&	=  e^{-\ii\bb\bl/2} \int_{\fM } \D\bq e^{ 2 \ii  \bq \bx}
	e^{\ii ( \bp - \bq +\bl) \bb} \delta^{[\bg]}(2\bq - \bl)  \\
\end{split}
\ee
here $\delta^\bg(\cdot)$ is a delta function modulo the reciprocal lattice vectors. To be more specific
\be
\delta^{[\bg]}(\bp) ={\det\{\belj\}} \prod_{j=1}^M \sum_{k_j\in\dZ} \delta\((\bp+k_j\bgj)\belj\)
={\det\{\belj\}}\prod_{j=1}^M \sum_{k_j\in\dZ} \delta(\bp \belj+2\pi k_j)
\ee
where ${\rm det}\, \{\belj\} = {\rm det}\, (...\belj...)$ is the determinant of matrix composed of vectors $\belj$. We took into account that $\bgj\belj=2\pi$.

Given that $\bq\in\fM$, there are the following non-vanishing contributions of $\delta^{[\bg]}(2\bq - \bl)$
\be
\bq = \bl/2, \quad \bl/2+\bg_{ U}/2
\ee
where $\bg_{ U} = \sum_{j\in U}(-1)^{\al_j}\bg^{(j)}$, $\al_j=0,1$
and $ U$ is any subset of $\{1,2,\ldots M\}$, i.e., for some particular choice of signs $\al_j$ depending on $\bl$, the combination $\bl/2+\bg_{ U}/2$ will also be inside $  \cM$ for any choice of $ U$. Thus we have
\be
\begin{split}
	Q_W
	&	=  \frac1{2^M}  e^{\ii (\bp \bb + \bl\bx) } \sum_{\forall  U } e^{-\ii\bg_{ U}\bb/2}.
\end{split}
\ee
We took into account that for any lattice vector $\bx\in\fO $  we have $\bg\bx = 2 \pi k$, $k\in\dZ$.  Now to have $Q_W(\bx,\bp) 	=    e^{\ii (\bp \bb + \bl\bx) }$ it is sufficient to require that $\bb$ is an even combination of the lattice basis vectors
\be
\bb = 2 \sum_{j=1}^M k_j \belj,\quad k_j\in\dZ.
\ee
Recalling that $e^{\ii \hat\bp \bb }$ is simply the translation operator in the coordinate representation, we come to the requirement that our class of operators must describe only the even jumps. Hence, we separate the original lattice into the even and the odd ones, $\fO=\cO\cup\cO'$, and associate the even one, $\cO$, with the physical crystal lattice.

\section{Discussion}
{
It is clear that in all real experiments on the Quantum Hall Effect the actual magnetic field is non-homogeneous (though, possibly, only slightly). Still, the quantization of the Hall conductivity follows the same rigorous rules as derived for the magnetic fields strictly constant over the sample. Up to now this apparent contradiction was not studied, and we presented here the first rigorous demonstration that in the non-homogeneous systems the Hall conductivity is given by a topological invariant.
}

To this end, we proposed the precise Wigner-Weyl calculus for lattice models of condensed matter physics or quantum field theory. We restricted ourselves to consideration of the models on rectangular lattices, but generalization is immediate to general triclinic ones. All operators in such models are defined on Hilbert space described in Sect. \ref{H-space}. The constructed Weyl symbols of these operators obey the properties listed in Sect. \ref{SectAxiom}. Our Weyl symbols differ from those proposed in the other papers, and allow us to derive the topological expression for the Hall conductivity presented in Sect. \ref{maincond}. This expression remains valid for arbitrary magnetic field unlike our previous considerations \cite{ZW2019,ZZ2019_2,FZ2019}, which were limited to magnetic fields much smaller than several thousands Tesla and of wavelengths much larger than 1 Angstrom. Thus, our present construction can be also used  for  consideration of artificial crystal lattices \cite{Jung2014,Scammell2019,Wang2019}, in which external magnetic field of order of several Tesla may be sufficient to observe the Hofstadter butterfly. The corresponding non-homogeneous systems can be investigated using the formalism proposed here.

The extension of the presented formalism may be relevant for the description of other non-dissipative transport phenomena (Chiral Separation Effect, Chiral Torsional Effect, Spin Hall Effect, etc), which were previously considered for homogeneous magnetic fields. With the aid of our precise Wigner-Weyl calculus the corresponding transport coefficients can be written in closed form, and their topological properties investigated.

\acknowledgments
Both authors are indebted for numerous discussions to M.~Suleymanov. I.V.F. is grateful for many discussions to V. Kupriaynov.
\appendix

\section{Appendix A. Wigner-Weyl calculus of Felix Buot}
\label{app:buot}

\subsection{$\cB$-symbol}

A version of precise Wigner-Weyl formalism for lattice models has been given long time ago by F. Buot \cite{Buot1974,Buot2009,Buot2013}. We present some excerpts here, for simplicity restricting ourselves to rectangular lattice $\cO$ with cubic Brillouin zone $\cM$.

In this formalism the Wigner transformation of a function $B(p_1,p_2)$ (where  $p_1, p_2 \in \cM$) is defined as follows:
\begin{equation}
\begin{aligned}
{B}_\CB(x,p) &\equiv \frac{1}{2^{D}}\int_{\fM}  \D{q} e^{\ii x q} B({p+q/2},{p-q/2}) \\
&= \int_{\cM}  \D{q} e^{2\ii x q} B({p+q},{p-q}).
\label{GWxAH02}
\end{aligned}\,.
\end{equation}
$\fM$ might be thought of as a doubled Brillouin zone: in the one-dimensional case for $\cM = (-\pi/2\ell,\pi/2\ell]$, $\fM = (-\pi/\ell,\pi/\ell]$, see Section \ref{H-space}. The symbol of an operator (which will be also called the $\cB$-symbol) is defined, correspondingly, as
\be
\begin{split}
	{B}_\CB(x,p)
	&	\equiv \int_{ {\cal M}} \D q e^{2\ii x q} \braket{ {p+q}| \hat{B} |{p-q}}\\
	&	
	= \sum_{z,y\in \cO} \int_{{\cal M}} \D q e^{2\ii x q} \braket{ {p+q}|z}\braket{ z| \hat{B}| y} \braket{ y |{p-q}} \\
	&
	= \frac{1}{|{\cal M}|}\sum_{z,y\in \cO} \int_{ {\cal M}} \D q e^{2\ii x q -\ii(p+q)z+\ii(p-q)y} \braket{ z| \hat{B}| y} \\
	&
	= \sum_{z,y\in \cO}  e^{-\ii p(z-y)} {\bf d}(2x - z-y) \braket{ z| \hat{B}| y},
	\label{GWxAH1}
\end{split}
\ee
where
$$
{\bf d}(w)
= \frac{1}{|{\cal M}|} \int_{\cM} \D{q} e^{\ii wq}.
$$
Here the integral is over the Brillouin zone. Function $B_\CB(x,p)$ is defined by \Ref{GWxAH1} for any real-valued $x$, not only for the values of $x\in \cO$. Function ${\bf d}(w)$ is reduced to $\delta_{w,0}$ for $w\in \cO$.
The inverse transformation reads
\begin{eqnarray}
\frac{1}{|\cM |}
\sum_{x \in \fO}e^{-\ii k x}Q_\cB(x,p) 		
& = &
\frac{1}{|\cM|}  \sum_{x \in \fO}\int_{\cM} \D{q} e^{2 \ii (q-k/2) x}
Q(p+q,p-q) \nonumber\\
&=&
2^D \int_{\cM} \D{q} \delta^{[2\pi/\ell]}(2q-k)
Q(p+q,p-q) \nonumber\\
&=&
Q(p+k/2,p-k/2).
\label{iQ0}
\end{eqnarray}
It gives
\begin{eqnarray}
Q(p,q) &=&
\frac{1}{|\cM|}\sum_{x \in \fO}e^{-\ii (p-q) x}Q_\cB(x,(p+q)/2) 	.	
\label{iQ0_}
\end{eqnarray}
Notice that the summation here goes over the more dense lattice $\fO$. Therefore, to restore the operator by its symbol we need  to know the values of the latter not only on the physical points  of $\cO$, but on the intermediate ones as well, $\cO'$.

It appears, that with this definition of the symbols the Moyal product expression is precise for $x \in \fO$:
\be
\label{Moyal22}
(\hat{A}\hat{B})_\CB(x,p)
= 	A_\CB(x,p)\star B_\CB(x,p) \equiv  A_\CB(x,p) e^{\frac{\ii}{2} \left( \cev{\partial}_{x}\vec{\partial_p}-\cev{\partial_p}\vec{\partial}_{x}\right )}B_\CB(x,p).
\ee
The proof is as follows
\begin{equation}
\begin{aligned}
(AB)_\CB(x,p )&=
\frac{1}{2^D} \int_{{\fM}} \D{P} \int_{\cal M} \D{R}
e^{\ii  x P}
\braket{p+{P}/{2} | \hat{A}  | R}
\braket{R | \hat{B}  | p-{P}/{2}}\\
&=\frac{1}{2^{2D}}\int_{{\fM}} \D P \int_{{\fM}}\D K
e^{\ii  x P}
\braket{p+{P}/{2} | \hat{A}  | p-{K}/{2}}
\braket{p-{K}/{2} | \hat{B}  | p-{P}/{2}}\\
&= \frac{2^D}{2^D\,2^{2D} }\int_{{\fM}} \D{q d k} e^{\ii  x (q+ k)}
\braket{p+{q}/{2}+{k}/{2} | \hat{A}  | p-{q}/{2}+{k}/{2}}
\braket{p-{q}/{2}+{k}/{2} | \hat{B}  | p-{q}/{2}-{k}/{2}}\\
&= \frac{1}{2^{2D}}\int_{{\fM}} \D{q d k}
\[  e^{\ii  x  q}
\braket{p+{q}/{2} | \hat{A}  | p-{q}/{2}}
\]
e^{\frac{k}{2}\cev{\partial}_p-\frac{q}{2}\vec{\partial}_p}
\[  e^{\ii  x   k}
\braket{p+{ k}/{2} |\hat{B}  | p-{ k}/{2}}
\]\\
&= \[\frac{1}{2^{D}} \int_{{\fM}} \D q  e^{\ii  x  q}
\braket{p+{q}/{2}| \hat{A}  | p-{q}/{2}}
\]
e^{\frac{\ii}{2} (\cev{\partial}_{x}\vec{\partial}_p - \cev{\partial}_p\vec{\partial}_x)}
\[\frac{1}{2^{D}} \int_{{\fM}} \D k e^{\ii  x   k}
\braket{p+{k}/{2} | \hat{B}  | p-{k}/{2}}
\].
\label{Z2}
\end{aligned}
\end{equation}
The factor ${2^{D}}$ in the third line results from the Jacobian. In addition, we took into account that the integration in the third line is over $q,k\in \fM$, that is the integration over $P=q+k, K=q-k$ is over the region of rhomboid form, that is contained inside $\fM$. However, due to the periodicity the integral over this figure is equal to $2^D$ times the integral over $P, K\in \fM$. This results in the factor $\frac{1}{2^D}$ in the third line.

It is similarly easy to check, that all of the properties \Ref{*-def}-\Ref{id-def} of {\it Definition} \ref{w-def} for this version of the Wigner-Weyl technique are satisfied, except for the last one, see following  subsections.

Certain properties of the symbol are exceptional. In particular,
\be
A_\cB (x+\ell,p) \star B_\cB (x,p)=0
\label{star-kernel}
\ee
for any two operators $\hat{A}$ and $\hat{B}$.
Let us check \Ref{star-kernel} directly, using the $x$-representation  \Ref{GWxAH} of the $\cB$-symbol
\be
\begin{split}
	A_\cB (x&+\ell,p) \star B_\cB (x,p)
	= \sum_{{z_{1,2},\bar z_{1,2} }\in \cO} \delta_{2(x+\ell),z_1+z_2} e^{-\ii p(z_1 - z_2)} \braket{ z_1|  \hat{A}|z_2} \star_{x,p} e^{-\ii p(\bar{z}_1-\bar{z}_2)}\delta_{2x,  \bar{z}_1+\bar{z}_2}\braket{ \bar{z}_1|  \hat{B}|\bar{z}_2}
	\\
	&
	= \sum_{{z_{1,2},\bar z_{1,2} }\in \cO} \delta_{2(x+\ell),z_1+z_2}
	e^{-\ii(p-\ii\vec\partial_x/2)(z_1 - z_2)} \braket{ z_1|  \hat{A}|z_2}
	e^{-\ii (p+\ii\cev\partial_x/2)(\bar{z}_1-\bar{z}_2)}
	\delta_{2x,  \bar{z}_1+\bar{z}_2}\braket{ \bar{z}_1|  \hat{B}|\bar{z}_2}
	\\
	&
	= \sum_{{z_{1,2},\bar z_{1,2} }\in \cO} \delta_{2x+2\ell+\bar{z}_1-\bar{z}_2,z_1+z_2}
	\delta_{2x-z_1 +z_2,  \bar{z}_1+\bar{z}_2}
	e^{-\ii p(z_1 - z_2)-\ii p(\bar{z}_1-\bar{z}_2)} \braket{ z_1|  \hat{A}|z_2}  \braket{ \bar{z}_1|  \hat{B}|\bar{z}_2}
	\label{A-1*B-proof}
\end{split}
\ee
valid for $x \in \fO$. The sum does not vanish if both Kronecker symbols are non zero, i.e. $z_1 - z_2 +   \bar{z}_1+\bar{z}_2 = 2\ell-\bar{z}_1+\bar{z}_2+z_1+z_2$. This leads to equation
\be
z_2 - \bar{z}_1 = \ell,
\ee
which is never satisfied for $z_2 , \bar{z}_1 \in \cO$, leading indeed to \Ref{star-kernel}.

\subsection{Star product without differentiation}

\label{SectBuoStar}
Let us represent the star product of $A_W(x,p)$ and $B_W(x,p)$ for $x \in \fO$ through the matrix elements of $\hat A$ and $\hat B$
	\begin{eqnarray}
	A_\CB(x,p) \star B_\CB(x,p)& = & \int_{\cal M} dk dq e^{\ii  2x k } \langle p+k|  \hat{A}|p-k\rangle \star e^{2\ii  x q }\langle p+q|  \hat{B}|p-q\rangle \nonumber\\
	& = & \frac{1}{|{\cal M}|^2}\sum_{z_1,z_2,\bar{z}_1,\bar{z}_2\in \cO}\int_{\cal M} dk dq e^{\ii  2x k +iz_1(p+k) - iz_2(p-k)} \langle z_1|  \hat{A}|z_2\rangle \star \nonumber\\&& e^{2\ii  x q +i\bar{z}_1(p+q) - i\bar{z}_2(p-q)}\langle \bar{z}_1|  \hat{B}|\bar{z}_2\rangle
	\nonumber\\
	& = & \frac{1}{|{\cal M}|^2} \sum_{z_1,z_2,\bar{z}_1,\bar{z}_2\in \cO}\int_{\cal M} dk dq e^{\ii  (2x+z_1+z_2) k +\ii p(z_1 - z_2)} \langle z_1|  \hat{A}|z_2\rangle \star \nonumber\\&& e^{\ii  (2x +\bar{z}_1+\bar{z}_2) q +\ii p(\bar{z}_1-\bar{z}_2) }\langle \bar{z}_1|  \hat{B}|\bar{z}_2\rangle
	\nonumber\\
	& = & \sum_{z_1,z_2,\bar{z}_1,\bar{z}_2\in \cO} \delta_{2x,-z_1-z_2} e^{ ip(z_1 - z_2)} \langle z_1|  \hat{A}|z_2\rangle \star \nonumber\\&& e^{\ii p(\bar{z}_1-\bar{z}_2)}\delta_{2x,  -\bar{z}_1-\bar{z}_2}\langle \bar{z}_1|  \hat{B}|\bar{z}_2\rangle
	\nonumber\\
	& = &  \sum_{z_1,z_2,\bar{z}_1,\bar{z}_2\in \cO} \delta_{2x,-z_1-z_2} e^{ i(p-i\vec{\partial_x}/2)(z_1 - z_2)} \langle z_1|  \hat{A}|z_2\rangle  \nonumber\\&& e^{\ii (p+i\cev{\partial_x}/2)(\bar{z}_1-\bar{z}_2)}\delta_{2x,  -\bar{z}_1-\bar{z}_2}\langle \bar{z}_1|  \hat{B}|\bar{z}_2\rangle
	\nonumber\\
	& = & \sum_{z_1,z_2,\bar{z}_1,\bar{z}_2\in \cO} \delta_{2x-\bar{z}_1+\bar{z}_2,-z_1-z_2} e^{ ip(z_1 - z_2)} \langle z_1|  \hat{A}|z_2\rangle  \nonumber\\&& e^{\ii p(\bar{z}_1-\bar{z}_2)}\delta_{2x + z_1 - z_2,  -\bar{z}_1-\bar{z}_2}\langle \bar{z}_1|  \hat{B}|\bar{z}_2\rangle
	\end{eqnarray}
	Next, using Eq. (\ref{GWxAH}) we obtain (for $x \in \fO$)
	\begin{eqnarray}
	A_\CB(x,p) \star B_\CB(x,p)
	& = & \sum_{2z,2\bar{z},u,\bar{u} \in \cO} \delta_{2x+\bar{u},z} \delta_{2x-{u},\bar{z}}\int \frac{\D{p}^\prime}{|{\cal M}|}\frac{d\bar{p}^\prime}{|{\cal M}|}e^{ ip^\prime u + i \bar{p}^\prime \bar{u}} A_\CB(z,p-p^\prime)A_\CB(\bar{z},p-\bar{p}^\prime)\label{starlattice}.
	\end{eqnarray}
One can see, that in order to define the star product of the Weyl symbols $A_\cB(x,p)$ and $B_\cB(x,p)$ for $x \in \fO$ we do not need to know the values of these functions for all real values of $x$. It is enough to know the values of the Weyl symbols for $x \in \fO$, which is the extension of the physical lattice (it is doubled in one-dimensional models, and enlarged $2^D$ times in general case).

\subsection{Trace and its properties}
There are two tentative definitions of the trace
\be
\Tr_\fO A_\cB
\equiv \sum_{x \in \fO} \int \frac{\D{p}}{|{\cal M}|}
A_\cB(x,p),
\qquad
\Tr_\cO A_\cB
\equiv \sum_{x \in \cO} \int \frac{\D{p}}{|{\cal M}|}
A_\cB(x,p),
\label{Tr_cB}
\ee
distinct in the summation set: $\fO$ in the former, and $\cO$ in the latter. Both satisfy \Ref{Tr-def-1}, \be
\Tr_\fO A_\cB = \Tr_\cO A_\cB= \tr\hat A,
\label{cB-Tr-def-1}
\ee
but only ``$\Tr_\fO$'' also cater for \Ref{Tr-def-2}
\be
\Tr_\fO (A_\cB \star B_\cB) = \Tr_\fO (A_\cB B_\cB).
\label{cB-Tr-def-2}
\ee
For the future notice, it is worth mentioning that
\be
\Tr_\fO (A_\cB(x+\ell,p))
= 	\Tr_\fO (A_\cB(x,p))
\label{Tr x+l}
\ee
since the shift of the argument of the symbol only results in the change of the initial point of the sum in $x\in\fO$. We obtain immediately
\be
\Tr_\fO\[A_\cB(x+\ell,p)\star B_\cB (x+\ell,p)\]
=\Tr_\fO\[A_\cB(x,p)\star B_\cB (x,p)\],
\ee
simply because
$
A_\cB(x+\ell,p)\star B_\cB (x+\ell,p)=  (\hat A \hat B)_\cB (x+\ell,p)$.
On the other hand,
\be
\begin{split}
	\Tr_\cO A_\cB(x+\ell,p)
	&
	= \frac{1}{|{\cM}|}\sum_{x\in \cO}\int_\cM \D{p}	
	A_\cB (x+\ell,p)\\
	&	
	= 	\frac{1}{|{\cM}|} \int_\cM \D{p}
	\sum_{x,z_{1,2}\in \cO} \delta_{2(x+\ell),z_1+z_2} e^{-\ii p(z_1 - z_2)} \braket{ z_1|  \hat{A}|z_2}
	=
	\sum_{x,z_{1,2}\in \cO} \delta_{2(x+\ell),z_1+z_2} \delta_{z_1, z_2} \braket{ z_1| \hat{A}|z_2}\\
	&		
	=0.
	\label{Tr_cO_ell}
\end{split}
\ee
And a similar property holds for a $\star$-product,
\be
\Tr_\cO \[A_\cB(x+\ell,p)\star B_\cB (x+\ell,p)\] =0.
\label{Tr_cO_Aell_Bell}
\ee

From the basic trace and $\star$ properties, \Ref{Moyal22} and \Ref{cB-Tr-def-1}, it can be trivially shown that
\be
\Tr_\cO A_\cB\star B_\cB=\Tr_\cO (AB)_\cB
= \Tr_\fO A_\cB\star B_\cB=\Tr_\fO (AB)_\cB= \tr \hat A\hat B.
\label{cB-Tr-A*B}
\ee
The trace without a star, on the other hand, gives
(with an arbitrary shift $\varkappa$ proportional to the distance between the adjacent points of the extended lattice $\fO$)
\be
\begin{split}
	\Tr_\fO [A_\cB (x&+\varkappa,p) B_\cB (x,p)]
	\equiv \frac{1}{|{\cM}|}\sum_{x \in \fO}\int_\cM \D{p}	A_\cB (x+\varkappa,p) B_\cB (x,p)
	\\
	&
	=\frac{1}{|{\cM}|}\sum_{x \in \fO}\int_\cM \D{p} \sum_{{z_{1,2},\bar z_{1,2} }\in \cO} \delta_{2(x+\varkappa),z_1+z_2} e^{-\ii p(z_1 - z_2)} \braket{ z_1|  \hat{A}|z_2} e^{-\ii p(\bar{z}_1-\bar{z}_2)}\delta_{2x, \bar{z}_1+\bar{z}_2}\braket{\bar{z}_1|  \hat{B}|\bar{z}_2}\\
	&
	=\sum_{{z_{1,2},\bar z_{1,2} }\in \cO} \delta_{2\varkappa+\bar{z}_1+\bar{z}_2,z_1+z_2}
	\delta_{z_1 - z_2,-\bar{z}_1+\bar{z}_2}
	\braket{ z_1|  \hat{A}|z_2}
	\braket{\bar{z}_1|  \hat{B}|\bar{z}_2}.
	\label{cB-Tr-Aell-B}
\end{split}
\ee
For the sum to be non-zero, it must hold, in particular, that
\be
2 z_2 =2 \varkappa +2\bar z_2,
\ee
which has solutions in $\cO$ only for  $\varkappa$ given by an even integer times $\ell$ (in every component, if we are dealing with a $D$-dimensional problem). For $\varkappa=0$ we get then
$$
\Tr_\fO [A_\cB (x,p) B_\cB (x,p)] = \tr \hat A \hat B,
$$
which together with \Ref{cB-Tr-A*B} proves that $\Tr_\fO$ of \Ref{Tr_cB} does satisfy \Ref{cB-Tr-def-2}
$$
\Tr_\fO [A_\cB (x,p) \star B_\cB (x,p)]
=	\Tr_\fO [A_\cB (x,p) B_\cB (x,p)].
$$

\subsection{$\cB$-symbol of unity and the other examples}

On the other hand, it is a straightforward calculation to show that \Ref{id-def} is not satisfied for $\cB$-symbol. In $1$D case (for simplicity) it reads,
\be
\begin{split}
	(\hat 1)_\cB
	&\equiv \int_{\cM}  \D{q} e^{2\ii x q} \braket{p+q|p-q}
	=\int_{-\pi/2\ell}^{\pi/2\ell}  \D{q} e^{2\ii x q} \delta^{[\pi/\ell]}(2q)\\
	&	=
	\frac12\int_{-\pi/2\ell}^{\pi/2\ell}  \D{q} e^{2\ii x q} \(\delta(q)+\frac{1}{2}\delta(q-\pi/2\ell) + \frac{1}{2}\delta(q+\pi/2\ell)\)\\	
	&	=\frac{1}{2}(1+{\rm cos}(\pi x/\ell)).
\end{split}
\ee
The Groenewold equation in its turn (for $\hQ\hG=\hat1$ and $x\in \fO$) becomes
\be
Q_\CB(x,p)\star G_\CB(x,p)=(\hat1)_\CB =
\begin{cases}
	1,& x\in\cO\\
	0,& x\in\cO'\\
\end{cases}.
\label{starexact0}
\ee
Thus, the Groenewold equation acquires the form, which is not convenient for the derivation of the proper expression for the Hall conductivity.

In a similar fashion we can consider another example: an exponential operator, corresponding to a jump between the adjacent points in $\cO$, $\hat{Q} = e^{2\ii p\ell}$,
\be
\begin{split}
	\(e^{2\ii p\ell}\)_\CB(x,p)
	&= \int_{-\pi/2\ell}^{\pi/2\ell} \D q e^{2\ii q x} e^{2\ii( p-  q)\ell}\delta^{[\pi/\ell]}(2q)\\
	&=
	\frac{1}{2}\int_{-\pi/2\ell}^{\pi/2\ell} \D q e^{2\ii q x} e^{2\ii(p-q)\ell} \(\delta(q)+\frac{1}{2}\delta(q-\pi/2\ell) + \frac{1}{2}\delta(q+\pi/2\ell)\) \\
	& =
	\frac{1}{2}\(e^{2\ii p \ell}  +\frac{1}{2}e^{2\ii (p-\pi/2+ \pi x/\ell)}
	+ \frac{1}{2}e^{2\ii (p+\pi/2- \pi x/\ell)}\)
	= \frac{1}{2}e^{2\ii p\ell}(1 -\cos( \pi x/\ell)).
\end{split}
\ee
One can see, that for $x\in \cO$ the $\cB$-symbol of this operator vanishes at all.
In a more general case of the homogeneous system with $\hat{Q} = f(p)$, where $f(p)$ is a function defined in momentum space with periodic boundary conditions, we have $Q_\CB(p,x) = \frac{1}{2}(f(p) +  f(p + \pi/2\ell){\rm cos}(\pi x/\ell))$. This property of the $\cB$-symbol of an operator does not allow to obtain slowly varying function of $x$ for $Q_\cB(x,p)$ in the case, when operator $\hat{Q}(\hat{x},\hat{p})$ depends on the operator $\hat x$ slowly. This does not allow to apply in practice the given version of Wigner-Weyl formalism to a wide class of lattice models. In particular, in the simplest tight-binding models the Dirac operator $Q(p)$ is proportional to $\cos 2p\ell$, and the corresponding $\cB$-symbol vanishes identically for $x\in \cO$.

\section{Naive extension of Wigner-Weyl calculus from continuum theory}
\label{app:conti_W}

Following the direct analogy with continuum theory one may define  Wigner transformations and Weyl symbols in lattice models with compact momentum space as follows.  Wigner transformation of a function $B(p_1,p_2)$ (where  $p_1, p_2 \in {\cal M}$) is defined as
\begin{equation} \begin{aligned}
{B}_\CC(x,p) \equiv \int_{\cM}  \D{q} e^{\ii x q} B({p+q/2},{p-q/2}) \label{GWxAH0}
\end{aligned}\,.
\end{equation}
Identifying $B(p_1,p_2)$ with the matrix elements of an operator $\hat B$, we come to the definition of the Weyl symbol of operator
(we denote it by $B_\CC$):
 \begin{equation} \begin{aligned}
{B}_\CC(x,p) \equiv \int_{\cM} \D{q} e^{\ii x q} \braket{ {p+q/2}| \hat{B} |{p-q/2}} \label{GWxAH}
\end{aligned}\,.
\end{equation}
Here integral is over the Brillouin zone. Moyal product of the two functions in phase space $f(x,p)$ and $g(x,p)$ is defined as 	$$f(x,p)\star g(x,p) \equiv f(x,p) e^{\frac{\ii}{2} \left( \cev{\partial}_{x}\vec\partial_p-\cev\partial_p\vec{\partial}_{x}\right )}g(x,p)$$
Let us consider the case, when operators $\hat A$ and $\hat B$ are almost diagonal, i.e. $\braket{ {p+q/2}| \hat{A} |{p-q/2}}$ and $\braket{ {p+q/2}| \hat{B} |{p-q/2}}$ are nonzero for arbitrary $p$ but small $q$  only (compared to the size of momentum space). At the level of the symbols of operators, it corresponds to the case when variation of $A_\CC(x,p)$ (and $B_\CC(x,p)$) as a function of $x$ may be neglected on the distances of the order of the lattice spacing. Below we assume that the considered operators satisfy this requirement. Then the following  expression follows
\begin{eqnarray}\label{Moyal2}
(\hat{A}\hat{B})_\CC(x,p) = 	A_\CC(x,p)\star B_\CC(x,p) = A_\CC(x,p) e^{\frac{\ii}{2} \left( \cev{\partial}_{x}\vec\partial_p-\cev\partial_p\vec\partial_{x}\right )}B_\CC(x,p)\,.
\end{eqnarray}
The proof is given in \cite{Suleymanov2019}. We repeat it here briefly
\begin{equation}\begin{aligned}
&(AB)_\CC(x,p )=
 \int_{{\cal M}} \D{P} \int_{\cal M} \D{R}
 	e^{\ii  x P}
 		\braket{p+{P}{2} | \hat{A}  | R}
		\braket{R |\hat{B} | p-{P}/{2}}\\
&=\frac{1}{2^D}\int_{{\cal M}} d P \int_{K/2 \in {\cal M}}dK
	e^{\ii  x P}
		\braket{p+{P}/{2}| \hat{A} | p-{K}/{2}}
		\braket{p-{K}/{2}|\hat{B} |p-{P}/{2}}\\
&= \frac{2^D}{2^D }\int_{{\cal M}} \D{ q d k} e^{\ii  x (q+ k)}
	\braket{p+{q}/{2}+{k}/{2}| \hat{A} |p-{q}/{2}+{k}/{2}}
	\braket{p-{q}/{2}+{k}/{2}|\hat{B} |p-{q}/{2}-{k}/{2}}\\
&= \int_{{\cal M}} \D{ q d k}
	\Big[  e^{\ii  x  q}
		\braket{p+{q}/{2}| \hat{A} |p-{q}/{2}}
	\Big]
	e^{\frac{k}{2}\cev{\partial}_p-\frac{q}{2}\vec{\partial}_p}
	\Big[  e^{\ii  x   k}
		\braket{p+{ k}/{2}|\hat{B} |p-{ k}/{2}}
	\Big]\\
&= \[ \int_{{\cal M}} \D{q}  e^{\ii  x  q}
		\braket{p+{q}/{2}| \hat{A} |p-{q}/{2}}
	\]
	e^{\frac{\ii}{2} (-\cev{\partial}_p\vec{\partial}_x+\cev{\partial}_{x}\vec{\partial}_p )}
	\[ \int_{{\cal M}} \D{k} e^{\ii  x   k}
		\braket{p+{k}/{2}|\hat{B} |p-{k}/{2}}
	\].
\label{Z}
\end{aligned}
\end{equation}
Here the bra- and ket- vectors in momentum space are defined modulo vectors of reciprocal lattice. In the second line we change variables
$$
	P = q+k , \quad K = q- k
$$
$$
	q = \frac{P+K}{2}, \quad  k =\frac{P-K}2
$$
with the Jacobian
$$
J = \left|\begin{array}{cc} 1 & 1 \\
-1 & 1 \end{array} \right| = 2^D.
$$
This results in the factor ${2^{D}}$ in the third line. Here $D$ is the dimension of space-time.

The transition between the second and the third lines of Eq. (\ref{Z}) is only approximate whenever the operators are not diagonal. If, however, the off-diagonal matrix elements are small, it becomes possible to substitute the region of the integration in $q$ and $k$ (that corresponds to $P,K/2 \in {\cal M}$) by ${\cal M}\otimes {\cal M}$.

This version of the Wigner-Weyl technique may be applied successfully to the lattice Dirac operator (the inverse Green function) and the Green function $G(p,q)$ itself (to be considered as matrix elements of an operator $\hat G$: $G(p+q/2,p-q/2) =\braket{ {p+q/2}| \hat{G} |{p-q/2}}$). Both are almost diagonal if the external electromagnetic field varies slowly, i.e. if its variation on the distance of the order of lattice spacing may be neglected. This occurs for the magnitudes of external magnetic field much smaller than thousands Tesla, and for the wavelengths much larger than $1$ Angstrom.
One has in this approximation
\begin{eqnarray}
	G_\CC(x,p)\star Q_\CC(x,p)=(\hat{G}\hat{Q})_\CC(x,p)&=&\int_{q\in {\cal M}} dq e^{\ii x q} \braket{ {p+q/2}| \hat{G} \hat{Q} |{p-q/2}} \nonumber
	\\
	&=&\int_{q \in {\cal M}} dq e^{\ii x q} \delta(q)=1\,.
\end{eqnarray}
Thus we have the Groenewold equation
\begin{eqnarray}
	Q_\CC(x,p)\star G_\CC(x,p)=1\,.
\end{eqnarray}
With this definition of Weyl symbols for $\hat{Q} = e^{2\ii p\ell}$ we have (in the one-dimensional model with the Brillouin zone $(-\pi,\pi]$):
$$
Q_\CC(x,p) = \int_{-\pi/2\ell}^{\pi/2\ell} dq e^{\ii q x} e^{\ii(p-q)\ell}\delta^{[\pi/\ell]}(q) = e^{2\ii p\ell}
$$
In a more general case of the homogeneous system with $\hat{Q} = f(p)$, where $f(p)$ is the function defined in momentum space with periodic boundary conditions, we have $Q_\CC(p,x) = f(p)$.

\section{Old-fashioned version of Wigner-Weyl formalism with the Weyl symbol defined by series}
\label{app:old_W}
An alternative version of Wigner-Weyl formalism has been proposed in \cite{Zubkov2016a,Zubkov2016b}, although still giving only approximate results for lattice models.

For the Green's function, the continuum symbol \Ref{GWxAH0} has been used,
$$
{G}_\CC(x,p)
	=\int_{\cM}  \D{q} e^{\ii x q} 	G({p+q/2},{p-q/2}).
$$
At the same time, for the operators defined by their functional symbols $\cQ(x,p)$,  ${\hat Q} = \cQ(\ii\partial_p,p)$, an implicit definition was proposed based on the integral equation
\be
\begin{split}
\int_{\fM} &\D{P d K}
	f(P,K) {Q}_\cW(-\ii \cev{\partial}_{K}+\ii\vec{\partial}_{P},{P}/2+{K}/{2})\, h(P,K)\\
& =
	\int_{\fM} \D{P d K} f(P,K) \cQ\Big(\ii \partial_{P}+ \ii\partial_{K},{P}/2+ {K}/{2}\Big) \, h({P},{K}),
\label{corrl}
\end{split}
\ee
which has to hold for arbitrary functions $f({P},{K})$ and $h({P},{K})$ defined on momentum space ${P},{K}\in \fM$.  Notice, that $\frac{\partial}{\partial ({P}/2+ {K}/{2})} = \partial_{P} + {\partial_{K}}$ and $\frac{\partial}{\partial ({P}/2 - {K}/{2})} = \partial_{P} - {\partial_{K}}$.

The practical calculation of $Q_W$ is as follows. We should represent
the integrand of the RHS in \Ref{corrl} in the following way (for that we use the commutation relations to rearrange the order of operators):
$$
\cQ\Big(\ii \partial_{P}+ \ii\partial_{K},{P}/2+ {K}/{2}\Big)
	= \sum_{i,j}(\ii\partial_{P})^i \mathfrak{q}_{ij}(P/2+K/2) (\ii\partial_{K})^j.
$$
If the functional symbol was given by its Taylor series around zero,
\be
	\cQ(x,p)=\sum_n x^n Q_n(p),
\ee
then
\be
\mathfrak{q}_{ab}(p)
	= \sum_{j=0}^{\infty} C_{a+b+j}^{a+j}C_{a+j}^j \[2^{-j} (\ii\partial_{p})^j Q_{a+b+j}(p)\].
\ee
The operator-valued function $\cQ$ is periodic in $P$ and $K$. However, this does not necessarily apply to the functions $\mathfrak{q}_{ij}$. Nevertheless, we have:
\be
{\cal Q}\Big(\ii \partial_{P}+ \ii\partial_{K},{P}/2+ {K}/{2}\Big)
	= \sum_{i,j} (\ii\partial_{P})^i \mathfrak{q}_{ij}([P/2+K/2]\,{\rm mod}\,g_j) (\ii\partial_{K})^j
\ee
where $g_j$ are vectors of the reciprocal lattice. Substituting this into the RHS of \Ref{corrl} we are able to perform the integration by parts, obtaining
\be
\begin{split}
	\int_{\fM} &\D{P d K}
	f(P,K) {Q}_\cW(-\ii \cev{\partial}_{K}+\ii\vec{\partial}_{P},{P}/2+{K}/{2})\, h(P,K)\\
	& =
	\int_{\fM} \D{P d K} f(P,K)  \(\sum_{i,j}
	(-\ii\cev\partial_{P})^i \mathfrak{q}_{ij}(P/2+K/2) (\ii\vec\partial_{K})^j \) \, h({P},{K}).
\end{split}
\ee
It is solved by
\be
{Q}_\cW\(x+y,p\)
	= \sum_{i,j} x^i \mathfrak{q}_{ij}(p) y^j,
\label{Q-cW-xy}
\ee
alternatively written as
$$
{Q}_\cW\(x,p\)
	= \sum_{i,j} \frac{1}{2^{i+j}}x^{i+j} \mathfrak{q}_{ij}(p)
	= \sum_{i} x^{i} \mathfrak{q}_{i0}(p).
$$
The later equality is true since it holds that
\be
\mathfrak{q}_{a0}(p)
	= \frac{1}{2^a}\sum_{b=0}^a \mathfrak{q}_{b,a-b}(p).
\ee

Let us consider the product
\begin{eqnarray}
r =	Q_\cW(x,p)\star G_\CC(x,p)
\end{eqnarray}
assuming that $\hG$ and $\hQ$ are inverse operators, $\hG\hQ=\hQ\hG=\hat 1$. We substitute to this expression the definition of $G_\CC$ and obtain:
\be
r
	=\int_\cM \D{P} \, Q_\cW(x+{\ii}\vec{\partial}_{p}/{2},{p}- {\ii}\vec{\partial}_x/{2}) e^{\ii  {P} x} G({p}+{P}/2,{p}-{P}/2).
\ee
In this expression the derivatives  $\vec{\partial}_{p}$ and $\vec{\partial}_x$ inside the arguments of $Q_\cW$ act only outside of this function, i.e. on $e^{\ii  {P} x} G({p}+{P}/2,{p}-{P}/2)$ and do not act inside the function $Q_\cW$, i.e. on $p$ and $x$ in its arguments. This gives
\be
r
	=\int_{\cal M} \D{P} e^{\ii  {P} x} Q_\cW(-\ii \cev{\partial}_{P}+{\ii}\vec{\partial}_{p}/{2},{p}+ {P}/{2}) G({p}+{P}/2,{p}-{P}/2).
	\label{app:r1}
\ee
Now let us suppose, that the function $G({p}+{P}/2,{p}-{P}/2)$ is nonzero only in the small vicinity of $P =0\bmod{\pi/\ell}$, i.e.
 $\hG$ is almost diagonal operator.
Then the integration by parts may be applied, although only approximately, which upon substituting $Q_\cW$ using \Ref{Q-cW-xy} gives
\be
\begin{split}
r
&	\approx
	\int_{\cM} \D{P} e^{\ii  {P} x} \(\sum_{i,j}
	(\ii\partial_{P})^i \mathfrak{q}_{ij}(p+P/2) (\ii\partial_{p}/2)^j \) G({p}+{P}/2,{p}-{P}/2)
	\\
&	=
	\int_{\cM} \D{P} e^{\ii  {P} x} \cQ(\ii {\partial}_{P}+ {\ii}{\partial}_{p}/{2},{p}+ {P}/{2})   G({p}+{P}/2,{p}-{P}/2)
=	\int_{\fM} \D{P} e^{\ii  {P} x} \delta^{{[\pi/\ell]}}(P)=1\quad
{\forall x\in\cO,}
\end{split}
\ee
and we come to the Groenewold equation, although only approximate.

Thus, in the same way, as for the case of $Q_\cC$, the given definition allows to deal effectively with the systems, where external fields vary slowly. The advantage of the formalism discussed in the current section is that the Weyl symbol may be calculated relatively easily for a certain class of operators $\hat{Q}$.

\bibliography{cross-ref,wigner3}

\begin{thebibliography}{86}
\expandafter\ifx\csname natexlab\endcsname\relax\def\natexlab#1{#1}\fi
\expandafter\ifx\csname bibnamefont\endcsname\relax
  \def\bibnamefont#1{#1}\fi
\expandafter\ifx\csname bibfnamefont\endcsname\relax
  \def\bibfnamefont#1{#1}\fi
\expandafter\ifx\csname citenamefont\endcsname\relax
  \def\citenamefont#1{#1}\fi
\expandafter\ifx\csname url\endcsname\relax
  \def\url#1{\texttt{#1}}\fi
\expandafter\ifx\csname urlprefix\endcsname\relax\def\urlprefix{URL }\fi
\providecommand{\bibinfo}[2]{#2}
\providecommand{\eprint}[2][]{\url{#2}}

\bibitem[{\citenamefont{Thouless et~al.}(1982)\citenamefont{Thouless, Kohmoto,
  Nightingale, and den Nijs}}]{Thouless1982}
\bibinfo{author}{\bibfnamefont{D.~J.} \bibnamefont{Thouless}},
  \bibinfo{author}{\bibfnamefont{M.}~\bibnamefont{Kohmoto}},
  \bibinfo{author}{\bibfnamefont{M.~P.} \bibnamefont{Nightingale}},
  \bibnamefont{and} \bibinfo{author}{\bibfnamefont{M.}~\bibnamefont{den Nijs}},
  \bibinfo{journal}{Phys. Rev. Lett.} \textbf{\bibinfo{volume}{49}},
  \bibinfo{pages}{405} (\bibinfo{year}{1982}).

\bibitem[{\citenamefont{Avron et~al.}(1983)\citenamefont{Avron, Seiler, and
  Simon}}]{Avron1983}
\bibinfo{author}{\bibfnamefont{J.~E.} \bibnamefont{Avron}},
  \bibinfo{author}{\bibfnamefont{R.}~\bibnamefont{Seiler}}, \bibnamefont{and}
  \bibinfo{author}{\bibfnamefont{B.}~\bibnamefont{Simon}},
  \bibinfo{journal}{Phys. Rev. Lett.} \textbf{\bibinfo{volume}{51}},
  \bibinfo{pages}{51} (\bibinfo{year}{1983}).

\bibitem[{\citenamefont{Fradkin}(1991)}]{Fradkin1991}
\bibinfo{author}{\bibfnamefont{E.}~\bibnamefont{Fradkin}},
  \emph{\bibinfo{title}{{Field Theories of Condensed Matter Physics}}}
  (\bibinfo{publisher}{Addison Wesley Publishing Company},
  \bibinfo{year}{1991}).

\bibitem[{\citenamefont{Hatsugai}(1997)}]{Hatsugai1997}
\bibinfo{author}{\bibfnamefont{Y.}~\bibnamefont{Hatsugai}},
  \bibinfo{journal}{J. Phys. Condens. Matter} \textbf{\bibinfo{volume}{9}},
  \bibinfo{pages}{2507} (\bibinfo{year}{1997}).

\bibitem[{\citenamefont{Qi et~al.}(2008)\citenamefont{Qi, Hughes, and
  Zhang}}]{Qi2008}
\bibinfo{author}{\bibfnamefont{X.-L.} \bibnamefont{Qi}},
  \bibinfo{author}{\bibfnamefont{T.~L.} \bibnamefont{Hughes}},
  \bibnamefont{and} \bibinfo{author}{\bibfnamefont{S.-C.} \bibnamefont{Zhang}},
  \bibinfo{journal}{Phys. Rev. B} \textbf{\bibinfo{volume}{78}},
  \bibinfo{pages}{195424} (\bibinfo{year}{2008}).

\bibitem[{\citenamefont{Kaufmann et~al.}(2016)\citenamefont{Kaufmann, Li, and
  Wehefritz-Kaufmann}}]{Kaufmann:2015lga}
\bibinfo{author}{\bibfnamefont{R.~M.} \bibnamefont{Kaufmann}},
  \bibinfo{author}{\bibfnamefont{D.}~\bibnamefont{Li}}, \bibnamefont{and}
  \bibinfo{author}{\bibfnamefont{B.}~\bibnamefont{Wehefritz-Kaufmann}},
  \bibinfo{journal}{Rev. Math. Phys.} \textbf{\bibinfo{volume}{28}},
  \bibinfo{pages}{1630003} (\bibinfo{year}{2016}), \eprint{1501.02874}.

\bibitem[{\citenamefont{Tong}(2016)}]{Tong:2016kpv}
\bibinfo{author}{\bibfnamefont{D.}~\bibnamefont{Tong}},
  \emph{\bibinfo{title}{Lectures on the quantum hall effect}}
  (\bibinfo{year}{2016}), \eprint{arXiv:1606.06687}.

\bibitem[{\citenamefont{Ishikawa and Matsuyama}(1986)}]{IshikawaMatsuyama1986}
\bibinfo{author}{\bibfnamefont{K.}~\bibnamefont{Ishikawa}} \bibnamefont{and}
  \bibinfo{author}{\bibfnamefont{T.}~\bibnamefont{Matsuyama}},
  \bibinfo{journal}{Z. Phys. C} \textbf{\bibinfo{volume}{33}},
  \bibinfo{pages}{41} (\bibinfo{year}{1986}).

\bibitem[{\citenamefont{Volovik}(1988)}]{Volovik1988}
\bibinfo{author}{\bibfnamefont{G.~E.} \bibnamefont{Volovik}},
  \bibinfo{journal}{JETP} \textbf{\bibinfo{volume}{67}}, \bibinfo{pages}{9}
  (\bibinfo{year}{1988}), \bibinfo{note}{zhETF, Vol. 94, No. 3(9), 123}.

\bibitem[{\citenamefont{Volovik}(2003)}]{Volovik2003a}
\bibinfo{author}{\bibfnamefont{G.~E.} \bibnamefont{Volovik}},
  \emph{\bibinfo{title}{{The Universe in a Helium Droplet}}}
  (\bibinfo{publisher}{Clarendon Press}, \bibinfo{address}{Oxford},
  \bibinfo{year}{2003}).

\bibitem[{\citenamefont{Matsuyama}(1987)}]{Matsuyama1987a}
\bibinfo{author}{\bibfnamefont{T.}~\bibnamefont{Matsuyama}},
  \bibinfo{journal}{Prog. Theor. Phys} \textbf{\bibinfo{volume}{77}},
  \bibinfo{pages}{711} (\bibinfo{year}{1987}).

\bibitem[{\citenamefont{Hasan and Kane}(2010)}]{HasanKane2010}
\bibinfo{author}{\bibfnamefont{M.~Z.} \bibnamefont{Hasan}} \bibnamefont{and}
  \bibinfo{author}{\bibfnamefont{C.~L.} \bibnamefont{Kane}},
  \bibinfo{journal}{Rev. Mod. Phys.} \textbf{\bibinfo{volume}{82}},
  \bibinfo{pages}{3045} (\bibinfo{year}{2010}).

\bibitem[{\citenamefont{Qi and Zhang}(2011)}]{Xiao-LiangQi2011}
\bibinfo{author}{\bibfnamefont{X.-L.} \bibnamefont{Qi}} \bibnamefont{and}
  \bibinfo{author}{\bibfnamefont{S.-C.} \bibnamefont{Zhang}},
  \bibinfo{journal}{Rev. Mod. Phys} \textbf{\bibinfo{volume}{83}},
  \bibinfo{pages}{1057} (\bibinfo{year}{2011}).

\bibitem[{\citenamefont{Volovik}(2011)}]{Volovik2011}
\bibinfo{author}{\bibfnamefont{G.~E.} \bibnamefont{Volovik}},
  \emph{\bibinfo{title}{{Topology of quantum vacuum}}} (\bibinfo{year}{2011}),
  \eprint{arXiv:1111.4627}.

\bibitem[{\citenamefont{Volovik}(2007)}]{Volovik2007}
\bibinfo{author}{\bibfnamefont{G.~E.} \bibnamefont{Volovik}}, in
  \emph{\bibinfo{booktitle}{Springer Lecture Notes in Physics 718/2007}},
  edited by \bibinfo{editor}{\bibfnamefont{W.~G.} \bibnamefont{Unruh}}
  \bibnamefont{and}
  \bibinfo{editor}{\bibfnamefont{R.}~\bibnamefont{Schutzhold}}
  (\bibinfo{publisher}{Springer}, \bibinfo{year}{2007}), pp.
  \bibinfo{pages}{31--73}, \eprint{cond-mat/0601372}.

\bibitem[{\citenamefont{Volovik}(2010)}]{VolovikSemimetal}
\bibinfo{author}{\bibfnamefont{G.~E.} \bibnamefont{Volovik}},
  \bibinfo{journal}{JETP Lett.} \textbf{\bibinfo{volume}{91}},
  \bibinfo{pages}{55} (\bibinfo{year}{2010}), \eprint{0912.0502}.

\bibitem[{\citenamefont{Gurarie}(2011)}]{Gurarie2011a}
\bibinfo{author}{\bibfnamefont{V.}~\bibnamefont{Gurarie}},
  \bibinfo{journal}{Phys. Rev. B} \textbf{\bibinfo{volume}{83}},
  \bibinfo{pages}{085426} (\bibinfo{year}{2011}).

\bibitem[{\citenamefont{Essin and Gurarie}(2011)}]{EssinGurarie2011}
\bibinfo{author}{\bibfnamefont{A.~M.} \bibnamefont{Essin}} \bibnamefont{and}
  \bibinfo{author}{\bibfnamefont{V.}~\bibnamefont{Gurarie}},
  \bibinfo{journal}{Phys. Rev. B} \textbf{\bibinfo{volume}{84}},
  \bibinfo{pages}{125132} (\bibinfo{year}{2011}).

\bibitem[{\citenamefont{Volovik}()}]{Volovik2016}
\bibinfo{author}{\bibfnamefont{G.~E.} \bibnamefont{Volovik}},
  \emph{\bibinfo{title}{{Topological Superfluids}}}, \eprint{arXiv:1602.02595}.

\bibitem[{\citenamefont{Nielsen and
  Ninomiya}(1981{\natexlab{a}})}]{NielsenNinomiya1981a}
\bibinfo{author}{\bibfnamefont{H.}~\bibnamefont{Nielsen}} \bibnamefont{and}
  \bibinfo{author}{\bibfnamefont{M.}~\bibnamefont{Ninomiya}},
  \bibinfo{journal}{Nuclear Physics B} \textbf{\bibinfo{volume}{193}},
  \bibinfo{pages}{173 } (\bibinfo{year}{1981}{\natexlab{a}}), ISSN
  \bibinfo{issn}{0550-3213}.

\bibitem[{\citenamefont{Nielsen and
  Ninomiya}(1981{\natexlab{b}})}]{NielsenNinomiya1981b}
\bibinfo{author}{\bibfnamefont{H.}~\bibnamefont{Nielsen}} \bibnamefont{and}
  \bibinfo{author}{\bibfnamefont{M.}~\bibnamefont{Ninomiya}},
  \bibinfo{journal}{Nuclear Physics B} \textbf{\bibinfo{volume}{185}},
  \bibinfo{pages}{20 } (\bibinfo{year}{1981}{\natexlab{b}}), ISSN
  \bibinfo{issn}{0550-3213}.

\bibitem[{\citenamefont{So}(1985)}]{So1985a}
\bibinfo{author}{\bibfnamefont{H.}~\bibnamefont{So}}, \bibinfo{journal}{Prog.
  Theor. Phys} \textbf{\bibinfo{volume}{74}}, \bibinfo{pages}{585}
  (\bibinfo{year}{1985}).

\bibitem[{\citenamefont{Kaplan}(1992)}]{Kaplan1992a}
\bibinfo{author}{\bibfnamefont{D.~B.} \bibnamefont{Kaplan}},
  \bibinfo{journal}{Phys. Lett. B} \textbf{\bibinfo{volume}{288}},
  \bibinfo{pages}{342} (\bibinfo{year}{1992}), \eprint{hep-lat/9206013}.

\bibitem[{\citenamefont{Golterman et~al.}(1993)\citenamefont{Golterman, Jansen,
  and Kaplan}}]{Golterman1993}
\bibinfo{author}{\bibfnamefont{M.~F.~L.} \bibnamefont{Golterman}},
  \bibinfo{author}{\bibfnamefont{K.}~\bibnamefont{Jansen}}, \bibnamefont{and}
  \bibinfo{author}{\bibfnamefont{D.~B.} \bibnamefont{Kaplan}},
  \bibinfo{journal}{Phys. Lett. B} \textbf{\bibinfo{volume}{301}},
  \bibinfo{pages}{219} (\bibinfo{year}{1993}), \eprint{hep-lat/9209003}.

\bibitem[{\citenamefont{Ho{\v r}ava}(2005)}]{Hovrava2005}
\bibinfo{author}{\bibfnamefont{P.}~\bibnamefont{Ho{\v r}ava}},
  \bibinfo{journal}{Phys. Rev. Lett.} \textbf{\bibinfo{volume}{95}},
  \bibinfo{pages}{016405} (\bibinfo{year}{2005}).

\bibitem[{\citenamefont{Creutz}(2008)}]{Creutz2008a}
\bibinfo{author}{\bibfnamefont{M.}~\bibnamefont{Creutz}},
  \bibinfo{journal}{JETP} \textbf{\bibinfo{volume}{2008}}, \bibinfo{pages}{017}
  (\bibinfo{year}{2008}).

\bibitem[{\citenamefont{Kaplan and Sun}(2012)}]{Kaplan2011}
\bibinfo{author}{\bibfnamefont{D.~B.} \bibnamefont{Kaplan}} \bibnamefont{and}
  \bibinfo{author}{\bibfnamefont{S.}~\bibnamefont{Sun}},
  \bibinfo{journal}{Phys. Rev. Lett.} \textbf{\bibinfo{volume}{108}},
  \bibinfo{pages}{181807} (\bibinfo{year}{2012}).

\bibitem[{\citenamefont{Coleman and Hill}(1985)}]{ColemanHill1985}
\bibinfo{author}{\bibfnamefont{S.}~\bibnamefont{Coleman}} \bibnamefont{and}
  \bibinfo{author}{\bibfnamefont{B.}~\bibnamefont{Hill}},
  \bibinfo{journal}{Phys. Lett. B} \textbf{\bibinfo{volume}{159}},
  \bibinfo{pages}{184} (\bibinfo{year}{1985}).

\bibitem[{\citenamefont{Lee}(1986)}]{Lee1986}
\bibinfo{author}{\bibfnamefont{T.}~\bibnamefont{Lee}}, \bibinfo{journal}{Phys.
  Lett. B} \textbf{\bibinfo{volume}{171}}, \bibinfo{pages}{247}
  (\bibinfo{year}{1986}).

\bibitem[{\citenamefont{Zhang and Zubkov}(2019{\natexlab{a}})}]{ZZ2019}
\bibinfo{author}{\bibfnamefont{C.~X.} \bibnamefont{Zhang}} \bibnamefont{and}
  \bibinfo{author}{\bibfnamefont{M.~A.} \bibnamefont{Zubkov}},
  \emph{\bibinfo{title}{{Influence of interactions on the anomalous quantum
  Hall effect}}} (\bibinfo{year}{2019}{\natexlab{a}}),
  \eprint{arXiv:1902.06545}.

\bibitem[{\citenamefont{Kubo et~al.}(1959)\citenamefont{Kubo, Hasegawa, and
  Hashitsume}}]{KuboHasegawa1959}
\bibinfo{author}{\bibfnamefont{R.}~\bibnamefont{Kubo}},
  \bibinfo{author}{\bibfnamefont{H.}~\bibnamefont{Hasegawa}}, \bibnamefont{and}
  \bibinfo{author}{\bibfnamefont{N.}~\bibnamefont{Hashitsume}},
  \bibinfo{journal}{Journal of the Physical Society of Japan}
  \textbf{\bibinfo{volume}{14}}, \bibinfo{pages}{56} (\bibinfo{year}{1959}).

\bibitem[{\citenamefont{Niu et~al.}(1985)\citenamefont{Niu, Thouless, and
  Wu}}]{Niu1985a}
\bibinfo{author}{\bibfnamefont{Q.}~\bibnamefont{Niu}},
  \bibinfo{author}{\bibfnamefont{D.~J.} \bibnamefont{Thouless}},
  \bibnamefont{and} \bibinfo{author}{\bibfnamefont{Y.-S.} \bibnamefont{Wu}},
  \bibinfo{journal}{Phys. Rev. B} \textbf{\bibinfo{volume}{31}},
  \bibinfo{pages}{3372} (\bibinfo{year}{1985}).

\bibitem[{\citenamefont{Altshuler et~al.}(1980)\citenamefont{Altshuler,
  Khmel'nitzkii, Larkin, and Lee}}]{Altshuler0}
\bibinfo{author}{\bibfnamefont{B.~L.} \bibnamefont{Altshuler}},
  \bibinfo{author}{\bibfnamefont{D.}~\bibnamefont{Khmel'nitzkii}},
  \bibinfo{author}{\bibfnamefont{A.~I.} \bibnamefont{Larkin}},
  \bibnamefont{and} \bibinfo{author}{\bibfnamefont{P.~A.} \bibnamefont{Lee}},
  \bibinfo{journal}{Phys.Rev.B} p. \bibinfo{pages}{5142}
  (\bibinfo{year}{1980}).

\bibitem[{\citenamefont{Altshuler and Aronov}(1985)}]{Altshuler}
\bibinfo{author}{\bibfnamefont{B.~L.} \bibnamefont{Altshuler}}
  \bibnamefont{and} \bibinfo{author}{\bibfnamefont{A.~G.}
  \bibnamefont{Aronov}}, \emph{\bibinfo{title}{Electron-electron inter-action
  in disordered systems (Editors: A}} (\bibinfo{publisher}{L. Efros},
  \bibinfo{address}{M. Pollak, Elsevier, North Holland, Amsterdam},
  \bibinfo{year}{1985}).

\bibitem[{\citenamefont{Zubkov and Wu}(2019)}]{ZW2019}
\bibinfo{author}{\bibfnamefont{M.~A.} \bibnamefont{Zubkov}} \bibnamefont{and}
  \bibinfo{author}{\bibfnamefont{X.}~\bibnamefont{Wu}},
  \emph{\bibinfo{title}{{Topological invariant in terms of the Green functions
  for the Quantum Hall Effect in the presence of varying magnetic field}}}
  (\bibinfo{year}{2019}), \eprint{arXiv:1901.06661}.

\bibitem[{\citenamefont{Fialkovsky and Zubkov}()}]{FZ2019}
\bibinfo{author}{\bibfnamefont{I.~V.} \bibnamefont{Fialkovsky}}
  \bibnamefont{and} \bibinfo{author}{\bibfnamefont{M.~A.}
  \bibnamefont{Zubkov}}, \emph{\bibinfo{title}{{Elastic deformations and
  Wigner-Weyl formalism in graphene}}}, \eprint{1905.11097}.

\bibitem[{\citenamefont{Zhang and Zubkov}(2019{\natexlab{b}})}]{ZZ2019_2}
\bibinfo{author}{\bibfnamefont{C.~X.} \bibnamefont{Zhang}} \bibnamefont{and}
  \bibinfo{author}{\bibfnamefont{M.~A.} \bibnamefont{Zubkov}},
  \bibinfo{journal}{JETP letters}  (\bibinfo{year}{2019}{\natexlab{b}}),
  \eprint{arXiv:1908.04138}.

\bibitem[{\citenamefont{Suleymanov and Zubkov}(2019)}]{Suleymanov2019}
\bibinfo{author}{\bibfnamefont{M.}~\bibnamefont{Suleymanov}} \bibnamefont{and}
  \bibinfo{author}{\bibfnamefont{M.~A.} \bibnamefont{Zubkov}},
  \bibinfo{journal}{Nucl. Phys. B} \textbf{\bibinfo{volume}{938}},
  \bibinfo{pages}{171} (\bibinfo{year}{2019}), \eprint{1811.08233}.

\bibitem[{\citenamefont{Groenewold}(1946)}]{Groenewold1946}
\bibinfo{author}{\bibfnamefont{H.~J.} \bibnamefont{Groenewold}},
  \bibinfo{journal}{Physica} \textbf{\bibinfo{volume}{12}},
  \bibinfo{pages}{405} (\bibinfo{year}{1946}).

\bibitem[{\citenamefont{Moyal}(1949)}]{Moyal1949}
\bibinfo{author}{\bibfnamefont{J.~E.} \bibnamefont{Moyal}}, in
  \emph{\bibinfo{booktitle}{{Proceedings of the Philosophical Society, 45}}}
  (\bibinfo{year}{1949}), pp. \bibinfo{pages}{99--124}.

\bibitem[{\citenamefont{Weyl}(1927)}]{Weyl1927}
\bibinfo{author}{\bibfnamefont{H.}~\bibnamefont{Weyl}},
  \bibinfo{journal}{Zeitschrift fur Physik} \textbf{\bibinfo{volume}{46}},
  \bibinfo{pages}{1} (\bibinfo{year}{1927}).

\bibitem[{\citenamefont{Wigner}(1932)}]{Wigner1932}
\bibinfo{author}{\bibfnamefont{E.~P.} \bibnamefont{Wigner}},
  \bibinfo{journal}{Phys. Rev} \textbf{\bibinfo{volume}{40}},
  \bibinfo{pages}{749} (\bibinfo{year}{1932}).

\bibitem[{\citenamefont{Ali and Englis}(2005)}]{Ali2005}
\bibinfo{author}{\bibfnamefont{S.~T.} \bibnamefont{Ali}} \bibnamefont{and}
  \bibinfo{author}{\bibfnamefont{M.}~\bibnamefont{Englis}},
  \bibinfo{journal}{Rev. Math. Phys.} \textbf{\bibinfo{volume}{17}},
  \bibinfo{pages}{391} (\bibinfo{year}{2005}).

\bibitem[{\citenamefont{Berezin and Shubin}(1972)}]{Berezin1972}
\bibinfo{author}{\bibfnamefont{F.~A.} \bibnamefont{Berezin}} \bibnamefont{and}
  \bibinfo{author}{\bibfnamefont{M.~A.} \bibnamefont{Shubin}},
  p.~\bibinfo{pages}{21} (\bibinfo{year}{1972}).

\bibitem[{\citenamefont{Curtright and Zachos}(2012)}]{Curtright2012}
\bibinfo{author}{\bibfnamefont{T.~L.} \bibnamefont{Curtright}}
  \bibnamefont{and} \bibinfo{author}{\bibfnamefont{C.~K.}
  \bibnamefont{Zachos}}, \bibinfo{journal}{Asia Pacific Physics Newsletter}
  \textbf{\bibinfo{volume}{1}}, \bibinfo{pages}{37} (\bibinfo{year}{2012}),
  \eprint{1104.5269}.

\bibitem[{\citenamefont{Zachos et~al.}(2005)\citenamefont{Zachos, Fairlie, and
  Curtright}}]{Zachos2005}
\bibinfo{author}{\bibfnamefont{C.}~\bibnamefont{Zachos}},
  \bibinfo{author}{\bibfnamefont{D.}~\bibnamefont{Fairlie}}, \bibnamefont{and}
  \bibinfo{author}{\bibfnamefont{T.}~\bibnamefont{Curtright}},
  \emph{\bibinfo{title}{{Quantum Mechanics in Phase Space}}}
  (\bibinfo{publisher}{World Scientific}, \bibinfo{address}{Singapore},
  \bibinfo{year}{2005}), ISBN \bibinfo{isbn}{978-981-238-384-6}.

\bibitem[{\citenamefont{Cohen}(1966)}]{Cohen1966}
\bibinfo{author}{\bibfnamefont{L.}~\bibnamefont{Cohen}},
  \bibinfo{journal}{Journal of Mathematical Physics}
  \textbf{\bibinfo{volume}{7}}, \bibinfo{pages}{781} (\bibinfo{year}{1966}).

\bibitem[{\citenamefont{Agarwal and Wolf}(1970)}]{Agarwal1970}
\bibinfo{author}{\bibfnamefont{G.~S.} \bibnamefont{Agarwal}} \bibnamefont{and}
  \bibinfo{author}{\bibfnamefont{E.}~\bibnamefont{Wolf}},
  \bibinfo{journal}{Phys. Rev. D} \textbf{\bibinfo{volume}{2}},
  \bibinfo{pages}{2161} (\bibinfo{year}{1970}).

\bibitem[{\citenamefont{G.}(1963)}]{E.C.1963}
\bibinfo{author}{\bibfnamefont{E.~C.} \bibnamefont{G.}},
  \bibinfo{journal}{Phys. Rev. Lett.} \textbf{\bibinfo{volume}{10}},
  \bibinfo{pages}{277} (\bibinfo{year}{1963}).

\bibitem[{\citenamefont{Glauber}(1963)}]{Glauber1963}
\bibinfo{author}{\bibfnamefont{R.~J.} \bibnamefont{Glauber}},
  \bibinfo{journal}{Phys. Rev} \textbf{\bibinfo{volume}{131}},
  \bibinfo{pages}{2766} (\bibinfo{year}{1963}).

\bibitem[{\citenamefont{Husimi}(1940)}]{Husimi1940}
\bibinfo{author}{\bibfnamefont{K.}~\bibnamefont{Husimi}}, in
  \emph{\bibinfo{booktitle}{Proc. Phys. Math. Soc. Jpn. 22}}
  (\bibinfo{year}{1940}), pp. \bibinfo{pages}{264--314}.

\bibitem[{\citenamefont{Cahill and J.}(1969)}]{Cahill1969}
\bibinfo{author}{\bibfnamefont{K.~E.} \bibnamefont{Cahill}} \bibnamefont{and}
  \bibinfo{author}{\bibfnamefont{R.}~\bibnamefont{J.}}, \bibinfo{journal}{Phys.
  Rev.} \textbf{\bibinfo{volume}{177}}, \bibinfo{pages}{1882}
  (\bibinfo{year}{1969}).

\bibitem[{\citenamefont{Buot}(2009)}]{Buot2009}
\bibinfo{author}{\bibfnamefont{F.~A.} \bibnamefont{Buot}},
  \emph{\bibinfo{title}{Nonequilibrium Quantum Transport Physics in
  Nanosystems}} (\bibinfo{publisher}{World Scientific}, \bibinfo{year}{2009}).

\bibitem[{\citenamefont{Lorce and Pasquini}(2011)}]{Lorce2011}
\bibinfo{author}{\bibfnamefont{C.}~\bibnamefont{Lorce}} \bibnamefont{and}
  \bibinfo{author}{\bibfnamefont{B.}~\bibnamefont{Pasquini}},
  \bibinfo{journal}{Phys.Rev. D} \textbf{\bibinfo{volume}{84}},
  \bibinfo{pages}{014015} (\bibinfo{year}{2011}), \eprint{1106.0139}.

\bibitem[{\citenamefont{Elze et~al.}(1986)\citenamefont{Elze, Gyulassy, and
  Vasak}}]{Elze1986}
\bibinfo{author}{\bibfnamefont{H.~T.} \bibnamefont{Elze}},
  \bibinfo{author}{\bibfnamefont{M.}~\bibnamefont{Gyulassy}}, \bibnamefont{and}
  \bibinfo{author}{\bibfnamefont{D.}~\bibnamefont{Vasak}},
  \bibinfo{journal}{Nucl. Phys. B} \textbf{\bibinfo{volume}{706}},
  \bibinfo{pages}{276} (\bibinfo{year}{1986}).

\bibitem[{\citenamefont{Hebenstreit et~al.}(2010)\citenamefont{Hebenstreit,
  Alkofer, and Gies}}]{Hebenstreit2010}
\bibinfo{author}{\bibfnamefont{F.}~\bibnamefont{Hebenstreit}},
  \bibinfo{author}{\bibfnamefont{R.}~\bibnamefont{Alkofer}}, \bibnamefont{and}
  \bibinfo{author}{\bibfnamefont{H.}~\bibnamefont{Gies}},
  \bibinfo{journal}{Phys. Rev. D} \textbf{\bibinfo{volume}{82}},
  \bibinfo{pages}{105026} (\bibinfo{year}{2010}), \eprint{1007.1099}.

\bibitem[{\citenamefont{Calzetta et~al.}(1988)\citenamefont{Calzetta, Habib,
  and Hu}}]{Calzetta1988}
\bibinfo{author}{\bibfnamefont{E.}~\bibnamefont{Calzetta}},
  \bibinfo{author}{\bibfnamefont{S.}~\bibnamefont{Habib}}, \bibnamefont{and}
  \bibinfo{author}{\bibfnamefont{B.~L.} \bibnamefont{Hu}},
  \bibinfo{journal}{Phys. Rev. D} \textbf{\bibinfo{volume}{37}},
  \bibinfo{pages}{2901} (\bibinfo{year}{1988}).

\bibitem[{\citenamefont{Bastos et~al.}(2008)\citenamefont{Bastos, Bertolami,
  Dias, and Prata}}]{Bastos2008}
\bibinfo{author}{\bibfnamefont{C.}~\bibnamefont{Bastos}},
  \bibinfo{author}{\bibfnamefont{O.}~\bibnamefont{Bertolami}},
  \bibinfo{author}{\bibfnamefont{N.~C.} \bibnamefont{Dias}}, \bibnamefont{and}
  \bibinfo{author}{\bibfnamefont{J.~N.} \bibnamefont{Prata}},
  \bibinfo{journal}{J. Math. Phys.} \textbf{\bibinfo{volume}{49}},
  \bibinfo{pages}{072101} (\bibinfo{year}{2008}), \eprint{[hep-th/0611257]}.

\bibitem[{\citenamefont{Dayi and Kelleyane}(2002)}]{Dayi2002}
\bibinfo{author}{\bibfnamefont{O.~F.} \bibnamefont{Dayi}} \bibnamefont{and}
  \bibinfo{author}{\bibfnamefont{L.~T.} \bibnamefont{Kelleyane}},
  \bibinfo{journal}{Mod. Phys. Lett. A} \textbf{\bibinfo{volume}{17}},
  \bibinfo{pages}{1937} (\bibinfo{year}{2002}), \eprint{[hep-th/0202062]}.

\bibitem[{\citenamefont{Habib and Laflamme}(1990)}]{Habib1990}
\bibinfo{author}{\bibfnamefont{S.}~\bibnamefont{Habib}} \bibnamefont{and}
  \bibinfo{author}{\bibfnamefont{R.}~\bibnamefont{Laflamme}},
  \bibinfo{journal}{Phys. Rev. D} \textbf{\bibinfo{volume}{42}},
  \bibinfo{pages}{4056} (\bibinfo{year}{1990}).

\bibitem[{\citenamefont{Chapman and Heinz}(1994)}]{Chapman1994}
\bibinfo{author}{\bibfnamefont{S.}~\bibnamefont{Chapman}} \bibnamefont{and}
  \bibinfo{author}{\bibfnamefont{U.~W.} \bibnamefont{Heinz}},
  \bibinfo{journal}{Phys. Lett. B} \textbf{\bibinfo{volume}{340}},
  \bibinfo{pages}{250} (\bibinfo{year}{1994}), \eprint{[hep-ph/9407405]}.

\bibitem[{\citenamefont{Berry}(1977)}]{Berry1977}
\bibinfo{author}{\bibfnamefont{M.~V.} \bibnamefont{Berry}},
  \bibinfo{journal}{Phil. Trans. Roy. Soc. Lond. A}
  \textbf{\bibinfo{volume}{287}}, \bibinfo{pages}{0145} (\bibinfo{year}{1977}).

\bibitem[{\citenamefont{Schwinger}(1960)}]{Schwinger570}
\bibinfo{author}{\bibfnamefont{J.}~\bibnamefont{Schwinger}},
  \bibinfo{journal}{Proceedings of the National Academy of Sciences}
  \textbf{\bibinfo{volume}{46}}, \bibinfo{pages}{570} (\bibinfo{year}{1960}),
  ISSN \bibinfo{issn}{0027-8424}.

\bibitem[{\citenamefont{Buot}(1974)}]{Buot1974}
\bibinfo{author}{\bibfnamefont{F.~A.} \bibnamefont{Buot}},
  \bibinfo{journal}{Phys.Rev. B} \textbf{\bibinfo{volume}{10}},
  \bibinfo{pages}{3700} (\bibinfo{year}{1974}).

\bibitem[{\citenamefont{Buot}(2013)}]{Buot2013}
\bibinfo{author}{\bibfnamefont{F.~A.} \bibnamefont{Buot}},
  \bibinfo{journal}{Quantum Matter} \textbf{\bibinfo{volume}{2}},
  \bibinfo{pages}{247} (\bibinfo{year}{2013}).

\bibitem[{\citenamefont{Wootters}(1987)}]{WOOTTERS19871}
\bibinfo{author}{\bibfnamefont{W.~K.} \bibnamefont{Wootters}},
  \bibinfo{journal}{Annals of Physics} \textbf{\bibinfo{volume}{176}},
  \bibinfo{pages}{1 } (\bibinfo{year}{1987}), ISSN \bibinfo{issn}{0003-4916}.

\bibitem[{\citenamefont{Leonhardt}(1995)}]{Leonhardt1995}
\bibinfo{author}{\bibfnamefont{U.}~\bibnamefont{Leonhardt}},
  \bibinfo{journal}{Phys. Rev. Lett.} \textbf{\bibinfo{volume}{74}},
  \bibinfo{pages}{4101} (\bibinfo{year}{1995}).

\bibitem[{\citenamefont{Kasperkovitz and Peev}(1994)}]{KASPERKOVITZ199421}
\bibinfo{author}{\bibfnamefont{P.}~\bibnamefont{Kasperkovitz}}
  \bibnamefont{and} \bibinfo{author}{\bibfnamefont{M.}~\bibnamefont{Peev}},
  \bibinfo{journal}{Annals of Physics} \textbf{\bibinfo{volume}{230}},
  \bibinfo{pages}{21 } (\bibinfo{year}{1994}), ISSN \bibinfo{issn}{0003-4916}.

\bibitem[{\citenamefont{Ligabò}(2016)}]{Ligabo2016}
\bibinfo{author}{\bibfnamefont{M.}~\bibnamefont{Ligabò}},
  \bibinfo{journal}{Journal of Mathematical Physics}
  \textbf{\bibinfo{volume}{57}}, \bibinfo{pages}{082110}
  (\bibinfo{year}{2016}).

\bibitem[{\citenamefont{Bayen et~al.}(1978)\citenamefont{Bayen, Flato,
  Fronsdal, Lichnerowicz, and Sternheimer}}]{BAYEN197861}
\bibinfo{author}{\bibfnamefont{F.}~\bibnamefont{Bayen}},
  \bibinfo{author}{\bibfnamefont{M.}~\bibnamefont{Flato}},
  \bibinfo{author}{\bibfnamefont{C.}~\bibnamefont{Fronsdal}},
  \bibinfo{author}{\bibfnamefont{A.}~\bibnamefont{Lichnerowicz}},
  \bibnamefont{and}
  \bibinfo{author}{\bibfnamefont{D.}~\bibnamefont{Sternheimer}},
  \bibinfo{journal}{Annals of Physics} \textbf{\bibinfo{volume}{111}},
  \bibinfo{pages}{61 } (\bibinfo{year}{1978}), ISSN \bibinfo{issn}{0003-4916}.

\bibitem[{\citenamefont{Kontsevich}(2003)}]{Kontsevich2003}
\bibinfo{author}{\bibfnamefont{M.}~\bibnamefont{Kontsevich}},
  \bibinfo{journal}{Letters in Mathematical Physics}
  \textbf{\bibinfo{volume}{66}}, \bibinfo{pages}{157} (\bibinfo{year}{2003}),
  ISSN \bibinfo{issn}{1573-0530}.

\bibitem[{\citenamefont{Felder and Shoikhet}(2000)}]{Felder2000}
\bibinfo{author}{\bibfnamefont{G.}~\bibnamefont{Felder}} \bibnamefont{and}
  \bibinfo{author}{\bibfnamefont{B.}~\bibnamefont{Shoikhet}},
  \bibinfo{journal}{Letters in Mathematical Physics}
  \textbf{\bibinfo{volume}{53}}, \bibinfo{pages}{75} (\bibinfo{year}{2000}),
  ISSN \bibinfo{issn}{1573-0530}.

\bibitem[{\citenamefont{Kupriyanov and Vassilevich}(2008)}]{Kupriyanov2008}
\bibinfo{author}{\bibfnamefont{V.~G.} \bibnamefont{Kupriyanov}}
  \bibnamefont{and} \bibinfo{author}{\bibfnamefont{D.~V.}
  \bibnamefont{Vassilevich}}, \bibinfo{journal}{The European Physical Journal
  C} \textbf{\bibinfo{volume}{58}}, \bibinfo{pages}{627}
  (\bibinfo{year}{2008}), ISSN \bibinfo{issn}{1434-6052}.

\bibitem[{\citenamefont{Zubkov and Khaidukov}(2017)}]{Zubkov2017}
\bibinfo{author}{\bibfnamefont{M.~A.} \bibnamefont{Zubkov}} \bibnamefont{and}
  \bibinfo{author}{\bibfnamefont{Z.~V.} \bibnamefont{Khaidukov}},
  \bibinfo{journal}{JETP Lett.} \textbf{\bibinfo{volume}{106}},
  \bibinfo{pages}{166} (\bibinfo{year}{2017}), \bibinfo{note}{[Pisma Zh. Eksp.
  Teor. Fiz. {\bf 106} no.3, 166]}.

\bibitem[{\citenamefont{Chernodub and Zubkov}(2017)}]{Chernodub2017}
\bibinfo{author}{\bibfnamefont{M.~N.} \bibnamefont{Chernodub}}
  \bibnamefont{and} \bibinfo{author}{\bibfnamefont{M.~A.}
  \bibnamefont{Zubkov}}, \bibinfo{journal}{Phys. Rev. D}
  \textbf{\bibinfo{volume}{96}}, \bibinfo{pages}{056006}
  (\bibinfo{year}{2017}), \eprint{1703.06516}.

\bibitem[{\citenamefont{Khaidukov and Zubkov}(2017)}]{Khaidukov2017}
\bibinfo{author}{\bibfnamefont{Z.~V.} \bibnamefont{Khaidukov}}
  \bibnamefont{and} \bibinfo{author}{\bibfnamefont{M.~A.}
  \bibnamefont{Zubkov}}, \bibinfo{journal}{Phys. Rev. D}
  \textbf{\bibinfo{volume}{95}}, \bibinfo{pages}{074502}
  (\bibinfo{year}{2017}), \eprint{1701.03368}.

\bibitem[{\citenamefont{Zubkov}(2018)}]{Zubkov2018a}
\bibinfo{author}{\bibfnamefont{M.~A.} \bibnamefont{Zubkov}},
  \bibinfo{journal}{Annals Phys} \textbf{\bibinfo{volume}{393}},
  \bibinfo{pages}{264} (\bibinfo{year}{2018}), \eprint{1610.08041}.

\bibitem[{\citenamefont{Zubkov}(2016{\natexlab{a}})}]{Zubkov2016a}
\bibinfo{author}{\bibfnamefont{M.~A.} \bibnamefont{Zubkov}},
  \bibinfo{journal}{Phys. Rev. D} \textbf{\bibinfo{volume}{93}},
  \bibinfo{pages}{105036} (\bibinfo{year}{2016}{\natexlab{a}}),
  \eprint{1605.08724}.

\bibitem[{\citenamefont{Zubkov}(2016{\natexlab{b}})}]{Zubkov2016b}
\bibinfo{author}{\bibfnamefont{M.~A.} \bibnamefont{Zubkov}},
  \bibinfo{journal}{Annals Phys} \textbf{\bibinfo{volume}{373}},
  \bibinfo{pages}{298} (\bibinfo{year}{2016}{\natexlab{b}}),
  \eprint{1603.03665}.

\bibitem[{\citenamefont{Kharzeev}(2014)}]{Kharzeev2014}
\bibinfo{author}{\bibfnamefont{D.~E.} \bibnamefont{Kharzeev}},
  \bibinfo{journal}{Prog. Part. Nucl. Phys} \textbf{\bibinfo{volume}{75}},
  \bibinfo{pages}{133} (\bibinfo{year}{2014}), \eprint{1312.3348}.

\bibitem[{\citenamefont{Metlitski and Zhitnitsky}(2005)}]{Metlitski2005}
\bibinfo{author}{\bibfnamefont{M.~A.} \bibnamefont{Metlitski}}
  \bibnamefont{and} \bibinfo{author}{\bibfnamefont{A.~R.}
  \bibnamefont{Zhitnitsky}}, \bibinfo{journal}{Phys. Rev.}
  \textbf{\bibinfo{volume}{D 72}}, \bibinfo{pages}{045011}
  (\bibinfo{year}{2005}).

\bibitem[{\citenamefont{Chernodub}(2016)}]{Chernodub2016}
\bibinfo{author}{\bibfnamefont{M.~N.} \bibnamefont{Chernodub}},
  \bibinfo{journal}{Phys. Rev. Lett} \textbf{\bibinfo{volume}{117}},
  \bibinfo{pages}{141601} (\bibinfo{year}{2016}), \eprint{1603.07993}.

\bibitem[{\citenamefont{Rubel}(1956)}]{Rubel1956}
\bibinfo{author}{\bibfnamefont{L.~A.} \bibnamefont{Rubel}},
  \bibinfo{journal}{Trans. Amer. Math. Soc.} \textbf{\bibinfo{volume}{83}},
  \bibinfo{pages}{417} (\bibinfo{year}{1956}).

\bibitem[{\citenamefont{Jung et~al.}(2014)\citenamefont{Jung, Raoux, Qiao, and
  MacDonald}}]{Jung2014}
\bibinfo{author}{\bibfnamefont{J.}~\bibnamefont{Jung}},
  \bibinfo{author}{\bibfnamefont{A.}~\bibnamefont{Raoux}},
  \bibinfo{author}{\bibfnamefont{Z.}~\bibnamefont{Qiao}}, \bibnamefont{and}
  \bibinfo{author}{\bibfnamefont{A.~H.} \bibnamefont{MacDonald}},
  \bibinfo{journal}{Phys. Rev. B} \textbf{\bibinfo{volume}{89}},
  \bibinfo{pages}{205414} (\bibinfo{year}{2014}).

\bibitem[{\citenamefont{Scammell and Sushkov}(2019)}]{Scammell2019}
\bibinfo{author}{\bibfnamefont{H.~D.} \bibnamefont{Scammell}} \bibnamefont{and}
  \bibinfo{author}{\bibfnamefont{O.~P.} \bibnamefont{Sushkov}},
  \bibinfo{journal}{Phys. Rev. B} \textbf{\bibinfo{volume}{99}},
  \bibinfo{pages}{085419} (\bibinfo{year}{2019}).

\bibitem[{\citenamefont{Wang et~al.}(2019)\citenamefont{Wang, Zihlmann, Liu,
  Makk, Watanabe, Taniguchi, Baumgartner, and Schönenberger}}]{Wang2019}
\bibinfo{author}{\bibfnamefont{L.}~\bibnamefont{Wang}},
  \bibinfo{author}{\bibfnamefont{S.}~\bibnamefont{Zihlmann}},
  \bibinfo{author}{\bibfnamefont{M.-H.} \bibnamefont{Liu}},
  \bibinfo{author}{\bibfnamefont{P.}~\bibnamefont{Makk}},
  \bibinfo{author}{\bibfnamefont{K.}~\bibnamefont{Watanabe}},
  \bibinfo{author}{\bibfnamefont{T.}~\bibnamefont{Taniguchi}},
  \bibinfo{author}{\bibfnamefont{A.}~\bibnamefont{Baumgartner}},
  \bibnamefont{and}
  \bibinfo{author}{\bibfnamefont{C.}~\bibnamefont{Schönenberger}},
  \bibinfo{journal}{Nano Letters} \textbf{\bibinfo{volume}{19}},
  \bibinfo{pages}{2371} (\bibinfo{year}{2019}).

\end{thebibliography}

\end{document}